\documentclass[aps,10pt,nofootinbib,groupedaddress,preprintnumbers,superscriptaddress,twocolumn]{revtex4}
\usepackage[dvipdfmx]{graphicx}
\usepackage{amssymb,amsfonts,amsmath,bm,color,multirow,cases,empheq,hyperref}
\usepackage{mathrsfs}

\usepackage{enumerate}
\usepackage{ulem}
\usepackage{afterpage}
\hypersetup{%
 setpagesize=false,
 bookmarksnumbered=true,%
 bookmarksopen=true,%
 colorlinks=true,%
 linkcolor=blue,%
 citecolor=red}
\def\bs{b^{\mathrm{s}}}
\def\bSPO{b_{\mathrm{SPO}}}
\def\d{\mathrm{d}}
\def\D{\Delta}
\def\H{\mathrm{H}}

\def\pa{\partial}
\def\qSPO{q_{\mathrm{SPO}}}
\def\rh{r_{\mathrm{H}}}
\def\s{\sigma}
\def\S{\Sigma}

\def\vp{\varphi}
\def\<{\langle}
\def\>{\rangle}
\def\={\equiv}
\def\2{I\hspace{-.1em}I}
\def\3{I\hspace{-.1em}I\hspace{-.1em}I}
\def\4{I\hspace{-.1em}V}
\def\ii{i\hspace{-.1em}i}
\def\iii{i\hspace{-.1em}i\hspace{-.1em}i}
\def\iv{i\hspace{-.1em}v}
\def\vi{v\hspace{-.1em}i}
\def\vii{v\hspace{-.1em}i\hspace{-.1em}i}
\def\viii{v\hspace{-.1em}i\hspace{-.1em}i\hspace{-.1em}i}
\def\ix{i\hspace{-.1em}x}

\begin{document}
%%%%%%%%%%%%%%%%%%%%%%%%%%%%%%%%%%%%%%%%%%%%%

\title{Complete classification of photon escape in the Kerr black hole spacetime}

\author{Kota Ogasawara}
\email{kota@tap.scphys.kyoto-u.ac.jp}
\affiliation{Theoretical Astrophysics Group, Department of Physics, Kyoto University, Kyoto 606-8502, Japan}

\author{Takahisa Igata}
\email{igata@post.kek.jp}
\affiliation{KEK Theory Center, Institute of Particle and Nuclear Studies, High Energy Accelerator Research Organization, Tsukuba 305-0801, Japan}

\date{\today}
%\pacs{04.70.Bw,97.60.Lf}
\preprint{KUNS-2842}
\preprint{KEK-Cosmo-0265}
\preprint{KEK-TH-2271}

%%%%%%%%%%%%%%%%%%%%%%%%%%%%%%%%%%%%%%%%%%%%%

\begin{abstract}
We consider necessary and sufficient conditions for photons emitted from the vicinity of a Kerr black hole horizon to escape to infinity.
The radial equation of motion determines necessary conditions for photons to reach infinity, and the polar angle equation of motion further restricts the allowed region of photon motion. 
Unlike emission from the equatorial plane, the latter restrictions are crucial for photon escape when the initial polar angle of the emission point is arbitrary.
We provide a visualization tool to analyze these two conditions and demonstrate a procedure for revealing photon escape. 
Finally, we completely identify the two-dimensional impact parameter space in which photons can escape. 
\end{abstract}

\maketitle

%%%%%%%%%%%%%%%%%%%%%%%%%%%%%%%%%%%%%%%%%%%%%
%%%%%%%%%%%%%%%%%%%%%%%%%%%%%%%%%%%%%%%%%%%%%
\section{Introduction}
%%%%%%%%%%%%%%%%%%%%%%%%%%%%%%%%%%%%%%%%%%%%%

The observations of the M87 galactic center have revealed a bright ring structure and associated shadow, which were formed by the central supermassive object and surrounding light sources~\cite{Akiyama:2019cqa}.
This result strongly suggests that the central object is a black hole, but the possibility of an alternative to a black hole has not been dismissed~\cite{Cardoso:2019rvt}.
In general, the difference between a black hole and other objects is more pronounced in phenomena around the equivalent radius of the horizon.
Therefore, it is very important to detect signals coming from as close as the horizon radius of a central object to uniquely identify it. 
For detections in electromagnetic wave observations, photons must be able to escape from the vicinity of the center to an observation point and must be able to climb the gravitational potential within the observable frequency range.
The critical indicators for the observability of such near-horizon phenomena are the photon escape probability and the redshift factor.

The escape of photons from the vicinity of a black hole was first revealed by the pioneering work of Synge, who evaluated photon escape cones in the Schwarzschild spacetime~\cite{Synge:1966okc}.
He showed that 50\% of photons emitted from the photon sphere escape to infinity,
while the remaining 50\% are captured by the black hole.
Furthermore, in the limit where the emission point approaches the horizon, the opening angle of a photon escape cone is arbitrarily close to zero. 
In other words, the escape cone rapidly disappears as the emission point inside the photon sphere approaches the horizon, which implies that the observability of the vicinity of the horizon is extremely low.
This fact would be expected quite naturally given the nature of a black hole, form which nothing can escape.

Is this inconvenient truth for observational verification universal, even if a black hole is spinning? The answer is no.
It has recently been reported that the escape of photons from the vicinity of a fast-spinning black hole horizon can have a large escape probability, contrary to a naive expectation from the Schwarzschild case.%
\footnote{Escape cones in the Kerr and Kerr--de Sitter spacetimes were numerically analyzed in Refs.~\cite{Semerak:1996,Stuchlik:2018qyz}.}
This series of reports originate from the result~\cite{Ogasawara:2016yfk,Ogasawara:2019mir} that a uniform emitter at rest in a locally nonrotating frame arbitrarily approaches the extremal Kerr horizon and still allows about 29.16\% of photons to escape to infinity.
For an emitter in a circular orbit of an extremal Kerr black hole, it was found that about 54.64\% of photons escape to infinity in the limit where the orbital radius arbitrarily approaches the innermost stable circular orbit radius~\cite{Igata:2019hkz}, which coincides with the horizon radius. Furthermore, the analytical value of the probability was recently found using the near-horizon geometry of an extremal Kerr black hole~\cite{Gates:2020els}.
In this case, the relativistic boost due to the source's proper motion causes such an increase in the escape probability that the Doppler blueshift exceeds the gravitational redshift~\cite{Igata:2019hkz,Gates:2020sdh}. 
Similar phenomena have also been reported for various other black holes with extremal geometry~\cite{Zhang:2020pay}.

The unusual behavior of the escape probability in the (near-)extremal Kerr black hole spacetime is essentially based on the near-horizon throat geometry~\cite{Bardeen:1972}.
Because of the throat geometry in the extremal limit, the coordinate radius of the emission point coincides with the horizon radius, while the proper spatial distance between them remains large. 
This nontrivial property provides a nonzero escape probability even when the emission point is close enough to the horizon radius.
The most important implication is that the vicinity of the horizon in the near-extremal Kerr spacetime is sufficiently observable, unlike the naive predictions in the Schwarzschild spacetime.

The near-horizon geometry of an extremal Kerr black hole also induces characteristic behaviors for phenomena that occur in the off-equatorial plane. 
It is known that spherical photon orbits, one of the most characteristic photon orbits in the Kerr spacetime, accumulate at the horizon radius~\cite{Bardeen:1973tla,Igata:2019pgb} (the so-called horizon class).
Since this occurs only in the range $\theta_{2}\leq \theta\leq \pi-\theta_{2}$ [where $\theta_{2}=\arccos{(2\sqrt{3}-3)^{1/2}}\simeq 47.05{}^\circ$] for the extremal case, the photon escape probability, associated with the spherical photon orbits, must also behave nontrivially in this range.

This paper aims to completely classify the necessary and sufficient range of parameters for photons emitted from the vicinity of a Kerr black hole horizon to be reachable to infinity. 
The emission point is not limited to the equatorial plane (as assumed in previous works), but instead takes an arbitrary polar angle. 
In other words, our condition for photon escape is also applicable to off-equatorial emission phenomena near the horizon.

This paper is organized as follows. 
In Sec.~\ref{sec:2} we review the equations of photon motion (i.e., the null geodesic equations) in the Kerr black hole spacetime 
and identify the allowed range of variables for physical motion. 
In Sec.~\ref{sec:SPO} we review the spherical photon orbits, which characterize photon escape from the vicinity of a Kerr black hole horizon. 
In Sec.~\ref{sec:escapable} we clarify the necessary and sufficient conditions for photons to escape from the vicinity of the horizon to infinity by use of the allowed region of motion and the spherical photon orbits, and
we develop a method of visualizing escapable regions in a two-dimensional photon impact parameter space.
In Sec.~\ref{sec:criticalvalues} we divide the range of the polar angle into four parts, and introduce critical values of an impact parameter to explicitly specify the escapable regions. 
Using the visualization method and the critical values, we completely identify the parameter region for the extremal case in Sec.~\ref{sec:classification_ext} and for the subextremal case in Sec.~\ref{sec:classification_sub}.
Section~\ref{sec:summary} is devoted to discussions.
In this paper, we use units in which $c=1$ and $G=1$.

%%%%%%%%%%%%%%%%%%%%%%%%%%%%%%%%%%%%%%%%%%%%%
%%%%%%%%%%%%%%%%%%%%%%%%%%%%%%%%%%%%%%%%%%%%%
\section{General null geodesic motion in the Kerr black hole spacetime}
\label{sec:2}
%%%%%%%%%%%%%%%%%%%%%%%%%%%%%%%%%%%%%%%%%%%%%

The Kerr metric in the Boyer-Lindquist coordinates is given by
\begin{align}
g_{\mu\nu}\d x^\mu \d x^\nu
=&-\frac{\S\D}{A}\d t^2+\frac{\S}{\D}\d r^2+\S\d\theta^2 \nonumber\\
&+\frac{A}{\S}\sin^2\theta\left(\d\vp-\frac{2Mar}{A} \d t\right)^2,
\label{def:metric}
\end{align}
where
\begin{align}
\S&\=r^2+a^2\cos^2\theta,~~
\D\=r^2-2Mr+a^2, \nonumber\\
A&\=\left(r^2+a^2\right)^2-a^2\Delta\sin^2\theta.
\end{align}
The metric is parametrized by two parameters: the mass $M$ and spin $a$. We can assume $a\geq0$ without loss of generality. Throughout this paper, we only consider the parameter range of the black hole spacetime $0\leq a\leq M$.
Then the event horizon is located at $r=r_\H\= M+\sqrt{M^2-a^2}$, where $\D$ vanishes.
The spacetime is stationary and axisymmetric with two corresponding Killing vectors $\xi^a$ and $\psi^a$, where $\xi^a\pa_a=\pa_t$ and $\psi^a\pa_a=\pa_\vp$.
Furthermore, the spacetime has the Killing tensor $K_{ab}$ defined by~\cite{Walker:1970un}
\begin{align}
K_{ab}\=&\; \S^2 (\d\theta)_a (\d\theta)_b + \sin^2\theta\left[\left(r^2+a^2\right)(\d\vp)_a -a(\d t)_a\right] \nonumber\\
&\times \left[\left(r^2+a^2\right)(\d\vp)_b -a(\d t)_b\right] -a^2\cos^2\theta g_{ab}.
\end{align}
We adopt units in which $M=1$ in what follows.

We consider photon motion in the Kerr black hole spacetime.
Let $k^a$ be a tangent vector to null geodesics parametrized by an affine parameter $\lambda$.
According to the existence of $\xi^a$, $\psi^a$, and $K_{ab}$, a photon has three constants of motion~\cite{Carter:1963},
\begin{align}
E&\=-\xi^a k_a=-k_t, ~~ L\=\psi^a k_a=k_\vp, \nonumber\\
{\cal Q}&\= K_{ab} k^a k^b-\left(L-aE\right)^2,
\end{align}
where $E$, $L$, and $\cal Q$ denote the conserved energy, angular momentum, and Carter constant, respectively.
We introduce the dimensionless impact parameters
\begin{align}
b\=\frac{L}{E}, ~~ q\=\frac{{\cal Q}}{E^2},
\end{align}
where we have assumed that $E>0$ because we only focus on photons that escape to infinity.
Rescaling $k^a$ by $E$, such that, $k^a/E \to k^a$, we obtain the null geodesic equations
\begin{align}
k^t&=\dot{t}=\frac{1}{\S}\left[a\left(b-a\sin^2\theta\right)+\frac{r^2+a^2}{\Delta}\left(r^2+a^2-ab\right)\right],\\
k^r&=\dot{r}=\frac{\sigma_r}{\S}\sqrt{R},\label{k^r}\\
k^\theta&=\dot{\theta}=\frac{\sigma_\theta}{\S}\sqrt{\Theta}, \label{k^theta}\\
\label{eq:kvp}
k^\vp&=\dot{\vp}=\frac{1}{\S}\left[\frac{b}{\sin^2\theta}-a+\frac{a}{\Delta}\left(r^2+a^2-ab\right)\right],
\end{align}
where $\sigma_r\=\mathrm{sgn}(\dot{r})$, $\sigma_\theta\=\mathrm{sgn}(\dot{\theta})$, the dots denote derivatives with respect to $\lambda$, and
\begin{align}
R(r)&\=\left(r^2+a^2-ab\right)^2-\D\left[q+(b-a)^2\right],
\label{def:R}\\
\Theta(\theta)&\=q-\cos^2\theta\left(\frac{b^2}{\sin^2\theta}-a^2\right).
\label{def:Theta}
\end{align}
The allowed region for photon motion is given by $R\geq0$ and $\Theta\geq0$.
Since the Kerr geometry is reflection symmetric with respect to the equatorial plane $\theta=\pi/2$, 
we consider only the range $0\leq\theta\leq\pi/2$ in what follows.

Here we clarify the allowed parameter range restricted by $R\geq0$.
Solving $R=0$ for $b$, we obtain
\begin{align}
b=b_1(r;q)\=\frac{-2ar+\sqrt{r\D\left[r^3-q(r-2)\right]}}{r(r-2)},
\label{b_1}
\\
b=b_2(r;q)\=\frac{-2ar-\sqrt{r\D\left[r^3-q(r-2)\right]}}{r(r-2)},
\label{b_2}
\end{align}
where $b_2$ diverges at $r=2$.
Since we can rewrite Eq.~\eqref{def:R} using $b_i$ ($i=1,2$) as
\begin{align}
R=-r(r-2)(b-b_1)(b-b_2),
\end{align}
we find that the allowed parameter range of $b$ derived from $R\geq0$ is given by
\begin{align}
\begin{array}{cll}
b\leq b_1,~b_2\leq b & \mathrm{for} & \rh\leq r<2, \\[2mm]
b_2\leq b\leq b_1 & \mathrm{for} & r\geq 2.
\end{array}
\label{allowed_R}
\end{align}
Here and hereafter, we focus only on photon dynamics outside the horizon, $r>r_{\mathrm{H}}$. 
Furthermore, from now on we will not consider $b_2\leq b$ for $\rh\leq r< 2$ because this range is for a negative-energy photon, and such a photon cannot escape to infinity.

We also clarify the allowed parameter range restricted by $\Theta\geq 0.$
The non-negativity of $\Theta$ reads
\begin{align}
q \geq \cos^2\theta\left(\frac{b^2}{\sin^2\theta}-a^2\right).
\label{q_ineq_min}
\end{align}
For $\theta<\pi/2$, the right-hand side is a quadratic function in $b$, and its minimum value is $q=-a^2\cos^2\theta$ at $b=0$.
For $q<0$, the inequality~\eqref{q_ineq_min} leads to $b^2<a^2\sin^2\theta<a^2$, and hence we have $q<0<a^2-b^2$. This relation provides a necessary condition for negative $q$, 
\begin{align}
a^2-b^2-q>0,
\label{eq:qnegativecond}
\end{align}
which will be useful for later discussions.
The allowed parameter range of $b$ derived from $\Theta\geq0$ is given by
\begin{align}
|b| \leq B(\theta;q) \= \tan\theta\sqrt{q+a^2\cos^2\theta}.
\label{allowed_Theta}
\end{align}
For $\theta=\pi/2$, we have 
$q\geq0$, so that there is no restriction of $b$ derived from $\Theta\geq0$.

Thus, the allowed region for photon motion is given by the common region of Eqs.~\eqref{allowed_R} and \eqref{allowed_Theta}.
Since $b_i$ ($i=1,2$) and $B$ depend on $q$, the region where a photon can exist in 
fixed $r$ and $\theta$ is given as a two-dimensional parameter region of $(b,q)$.

%%%%%%%%%%%%%%%%%%%%%%%%%%%%%%%%%%%%%%%%%%%%%
\section{Spherical photon orbits}
\label{sec:SPO}
%%%%%%%%%%%%%%%%%%%%%%%%%%%%%%%%%%%%%%%%%%%%%

We review the spherical photon orbits, which characterize photon escape from the vicinity of the horizon~\cite{Ogasawara:2019mir}.
These are the orbits with $\dot{r}=0$ and $\ddot{r}=0$ outside the horizon~\cite{Teo:2003}. Through Eq.~\eqref{k^r}, they lead to
\begin{align}
R=0, \quad
\frac{\d R}{\d r}=0.
\label{R=R'=0}
\end{align}
Solving these coupled algebraic equations for $b$ and $q$, we obtain two sets of solutions.
One solution set is given by
\begin{align}
b=\frac{r^2+a^2}{a}, ~~ q=-\frac{r^4}{a^2}. \label{SPO_bq_unsuitable}
\end{align}
Since $q<0$ these parameters must satisfy the
condition~\eqref{eq:qnegativecond}, but they do not because $a^2-b^2-q=-2r^2<0$. 
Therefore, this solution set is unsuitable.

The other solution set is given by
\begin{align}
b&=b_{\mathrm{SPO}}(r) \= -\frac{r^3-3r^2+a^2r+a^2}{a(r-1)},
\label{bSPO}\\
q&=q_{\mathrm{SPO}}(r) \= -\frac{r^3 \left(r^3-6r^2+9r-4a^2\right)}{a^2\left(r-1\right)^2}. 
\label{qSPO}
\end{align}
Outside the horizon, $\qSPO$ has a unique local maximum with the value $27$ at $r=3$.
These parameters lead to
\begin{align}
a^2-b^2-q=-\frac{2r\left(r^3-3r+2a^2\right)}{(r-1)^2}<0
\label{Eq:SPO_q<0}
\end{align}
for $r>\rh$, and hence, the condition~\eqref{eq:qnegativecond} can never hold.%
\footnote{
Here we prove the inequality~\eqref{Eq:SPO_q<0}, i.e., we show $r^3-3r+2a^2 \geq 0$ for $r>\rh$.
Using the relations $(r-\rh)^3=r^3-3r^2\rh+3r\rh^2-\rh^3$, $\D(\rh)=\rh^2-2\rh+a^2=0$, and $1\leq\rh<r$, we have
\begin{align}
&r^3-3r+2a^2 \nonumber\\
&=\rh^3-3\rh+2a^2 + (r-\rh)\left[(r-\rh)^2+3(r\rh-1)\right] \nonumber\\
&>\rh^3-3\rh+2a^2=\left(1-a^2\right)\rh \geq 0.
\end{align}}
This means that, for the spherical photon orbit to exist, $q$ must satisfy the inequality $0 \leq q \leq 27$.

%%%%%%%%%%%%%%%%%%%%%%%%%%%%%%%%%%%%%%%%%%%%%
\begin{figure*}[t]
\centering
\includegraphics[width=12cm]{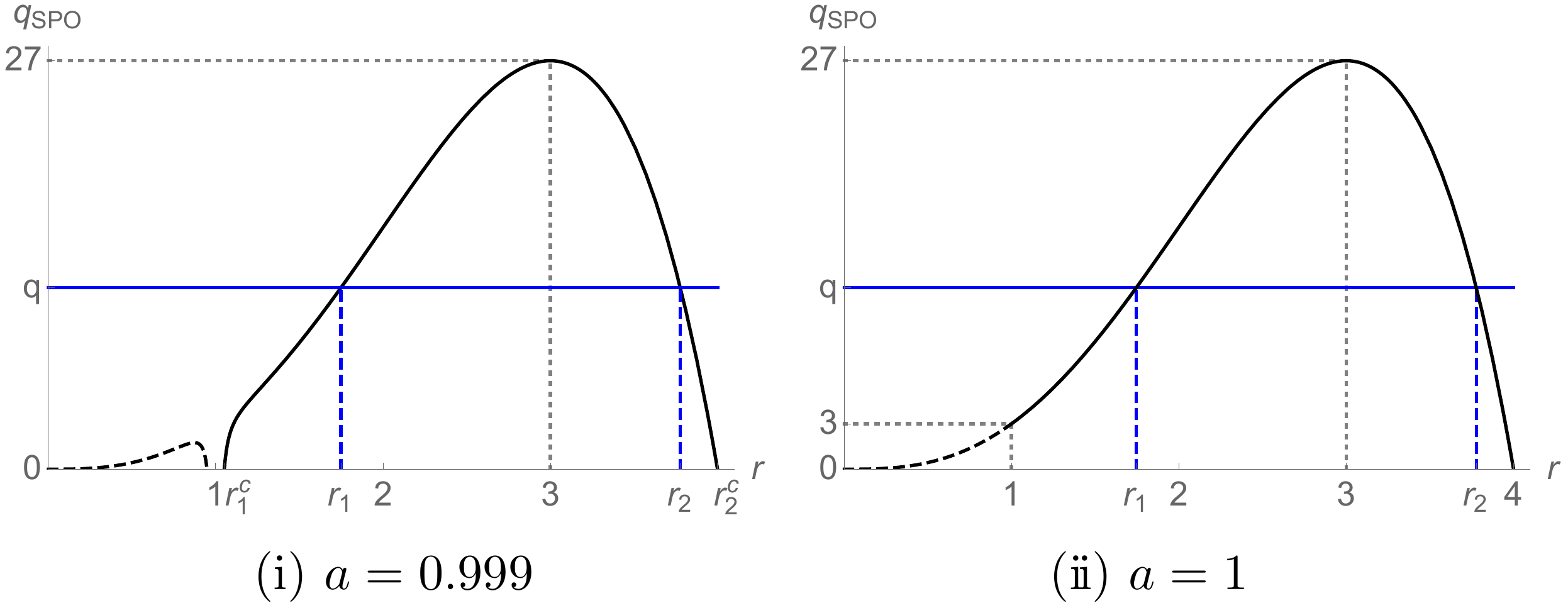}
\caption{
Relation between $q$ and the radii $r_i$ ($i=1,2$) of extremum points of $b_i$.
The function $\qSPO(r)$ is shown by black curves, which are solid outside the horizon and dashed inside it.
The intersections of the blue solid lines $q$ and the black solid curves $\qSPO(r)$ give the radii of spherical photon orbits $r_1$ and $r_2$.
The left panel (i) is the subextremal case ($a=0.999$), and the right panel (\ii) is the extremal case ($a=1$).}
\label{fig:qSPO}
\end{figure*}
%%%%%%%%%%%%%%%%%%%%%%%%%%%%%%%%%%%%%%%%%%%%%

First, we focus on the case of the subextremal Kerr black hole spacetime (i.e., $a<1$).
Figure~\ref{fig:qSPO}(i) shows a typical shape of $\qSPO(r)$.
Solving Eq.~\eqref{qSPO} for $r$, we obtain two roots $r=r_1(q)$ and $r=r_2(q)$ outside the horizon, which are the radii of spherical photon orbits.
Note that $r_1$ ($r_2$) increases (decreases) monotonically with $q$ in the range
\begin{align}
\rh<r^{\mathrm{c}}_1\leq r_1(q)\leq 3\leq r_2(q)\leq r^{\mathrm{c}}_2,
\label{ineq:r_i}
\end{align}
where
\begin{align}
r^{\mathrm{c}}_1&\=r_1(0)=2+2\cos\left[\frac{2}{3}\arccos(a)-\frac{2\pi}{3}\right],\\
r^{\mathrm{c}}_2&\=r_2(0)=2+2\cos\left[\frac{2}{3}\arccos(a)\right]
\end{align}
are the radii of circular photon orbits solving $\qSPO(r)=0$, i.e., $q=0$.
We obtain a direct relation between $b$ and $q$ for the spherical photon orbits by eliminating $r$ from Eqs.~\eqref{bSPO} and \eqref{qSPO}. Substituting $r=r_i(q)$ into $b_{\mathrm{SPO}}(r)$, we have
$b$ for the spherical photon orbits as a function of $q$, 
\begin{align}
\bs_i(q) \= \bSPO(r_i) ~~ (i=1,2),
\end{align}
which also correspond to the extremum values of $b_i(r;q)$. 
In the case $q=27$, $r_i$ coincide with $r=3$, so that 
$\bs_i$ coincide with $b=-2a$.
We also define the specific values $b^{\mathrm{c}}_i$ as $\bs_i(0)$, or, equivalently, 
as the value of $b$ for a photon in circular orbits,
\begin{align}
b^{\mathrm{c}}_i\=\bSPO(r^{\mathrm{c}}_i) ~~ (i=1,2).
\end{align}

Next, we focus on the extremal Kerr black hole spacetime (i.e., $a=1$).
In this case, Eqs.~\eqref{bSPO} and \eqref{qSPO} become
\begin{align}
b&=b_{\mathrm{SPO}}(r) = -r^2+2r+1, 
\label{bSPOextremal}\\
q&=q_{\mathrm{SPO}}(r) = r^3(4-r). \label{qSPOextremal}
\end{align}
Figure~\ref{fig:qSPO}(\ii) shows the shape of $\qSPO(r)$ for $a=1$.
Unlike the subextremal case, the number of roots of Eq.~\eqref{qSPOextremal} depends on $q$ outside the horizon.
There exists a single root $r_2$ for $0\leq q\leq3$, while there exist two roots $r_1$ and $r_2$ for $3<q\leq27$.
In the case $q=27$, $r_i$ coincide with $r=3$, so that $\bs_i$ coincide with $b=-2$.

%%%%%%%%%%%%%%%%%%%%%%%%%%%%%%%%%%%%%%%%%%%%%
\section{Escape conditions}
\label{sec:escapable}
%%%%%%%%%%%%%%%%%%%%%%%%%%%%%%%%%%%%%%%%%%%%%

We consider the conditions for photons escaping from the vicinity of the horizon to infinity, i.e., the escape conditions.
Let $(r_*, \theta_*)$ be the radial and polar angle coordinates of the emission point, respectively.
From the reflection symmetry of the background, 
we only consider $0\leq \theta_*\leq \pi/2$, in which $\theta_*=0, \pi/2$ will be considered separately in Appendix \ref{App:theta-0-pi}. The necessary and sufficient conditions for photons to escape are that they have appropriate parameters to reach infinity from $r=r_*$ and are in the allowed region determined by the variable $\theta_*$. 
In the following subsections, we consider the escape conditions for $q\geq 0$ and $q<0$ separately.

%%%%%
\subsection{$q\geq 0$}
\label{subsec:q>0}
%%%%%
Let us consider the behavior of $b_i(r;q)$ to determine the range of $b$ in which a photon with $q\geq0$ satisfies the necessary conditions to escape from $r=r_*$ to infinity~\cite{Ogasawara:2019mir}.
From now on, we consider the case where $r_*$ is in the range $\rh<r_*<3$.
There are three cases according to $r_*$ and the shape of $b_1(r;q)$, i.e., according to the relative position of $r_1$ to $\rh$ and $r_*$.

\textit{Case~(a)}---$r_1<\rh<r_*$:
The first inequality $r_1<\rh$ leads to $0\leq q<3$.
Note that the inequality $r_1<\rh$ appears only for $a=1$ 
[see Fig.~\ref{fig:qSPO}(\ii)].

\textit{Case~(b)}---$\rh\leq r_1<r_*$:
The corresponding range of $q$ is given by $3\leq q<q_*$ for $a=1$ and $0\leq q<q_*$ for $a<1$,
where
\begin{align}
q_*\=\qSPO(r_*).
\end{align}

\textit{Case~(c)}---$\rh<r_*\leq r_1$:
For $a=1$, the corresponding range of $q$ is given by $q_* \leq q \leq 27$.
For $a<1$, when $r_* \geq r_1^\mathrm{c}$, the corresponding range of $q$ is given by $q_* \leq q \leq 27$.
On the other hand, when $r_*<r_1^\mathrm{c}$, since $q_*<0$, the corresponding range of $q$ is given by $0 \leq q \leq 27$.

%%%%%%%%%%%%%%%%%%%%%%%%%%%%%%%%%%%%%%%%%%%%%
\begin{figure*}[t]
\centering
\includegraphics[width=16cm]{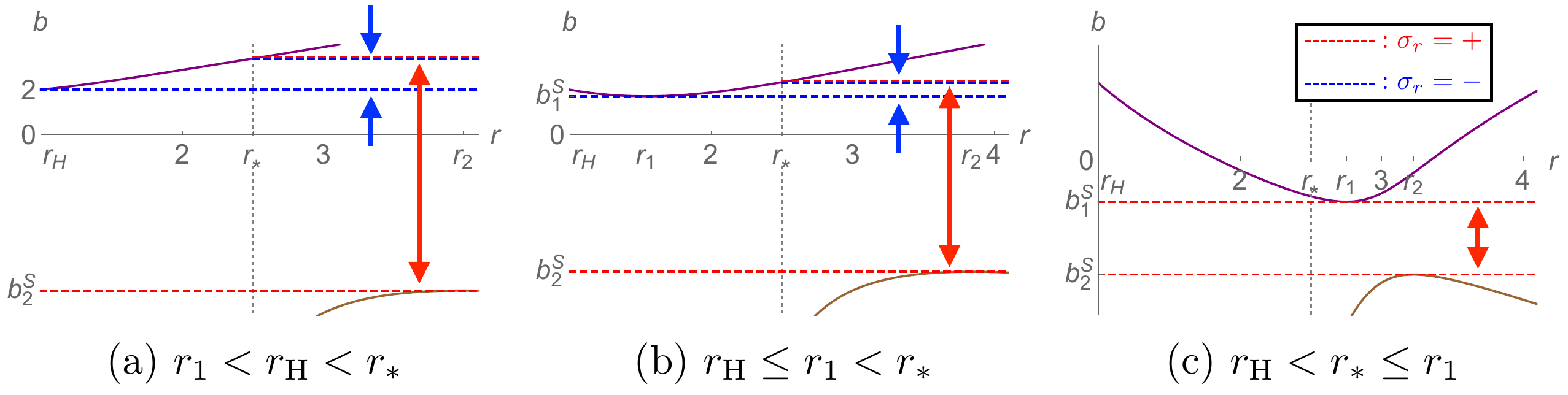}
\caption{
Typical shape of $b_i(r;q)$ in Cases~(a)--(c). 
The purple and brown solid curves
denote $b_1(r;q)$ and $b_2(r;q)$, respectively. Note that $b_2(r;q)$ in the range $\rh\leq r<2$ is not plotted.
The range of $b$ in which a photon satisfies a necessary condition for escape from $r=r_*$ to infinity depends on two conditions.
One is the relative position of $r_1$ to $\rh$ and $r_*$, and the other is whether a photon is emitted radially outward ($\s_r=+$) or inward ($\s_r=-$).
If a photon is emitted radially outward (inward), the maximum and minimum values of $b$ with which a photon satisfies the necessary condition for escape are given by the red (blue) dashed lines.
}
\label{fig:b3cases}
\end{figure*}
%%%%%%%%%%%%%%%%%%%%%%%%%%%%%%%%%%%%%%%%%%%%%

We can summarize all of the cases as follows:
\begin{widetext}
\begin{align}
\text{Case~(a):}~& r_1<\rh<r_* ~\Leftrightarrow~~ 0\leq q<3 ~~(\mathrm{for}~a=1),\\
\text{Case~(b):}~& \rh\leq r_1<r_* ~\Leftrightarrow ~ \left\{
\begin{array}{ll}
3\leq q <q_* & (\mathrm{for}~a=1), \\
0\leq q <q_* & (\mathrm{for}~a<1),
\end{array}
\right.\\
\text{Case~(c):}~& \rh<r_*\leq r_1 ~\Leftrightarrow~ \left\{
\begin{array}{ll}
q_* \leq q \leq 27 & (\mathrm{for}~a=1~ \mathrm{and~ for}~a<1~\mathrm{with}~r_* \geq r_1^\mathrm{c}), \\
0\leq q \leq 27 & (\mathrm{for}~a<1~\mathrm{with}~r_* < r_1^\mathrm{c}).
\end{array}
\right.
\end{align}
\end{widetext}
In other words, in the range $0\leq q\leq 27$,
only Case~(c) appears for $a<1$ and $r_*<r^{\mathrm{c}}_1$, 
and Cases~(b) and (c) appear for $a<1$ and $r_*\geq r^{\mathrm{c}}_1$, 
and all of the cases appear for $a=1$.

For Case~(a), as $r$ increases from $\rh$ to $\infty$, $b_1$ begins at $b_1(\rh;q)=2$ and monotonically increases to $\infty$.
For Cases~(b) and (c), as $r$ increases from $\rh$ to $\infty$, $b_1$ monotonically decreases from $b_1(\rh;q)=2\rh/a$ to a local minimum $\bs_1$ at $r=r_1$ and monotonically increases from there to $\infty$.
For all of the cases, as $r$ increases from $\rh$ to $2$, $b_2$ begins at $b_2(\rh;q)=b_1(\rh;q)$ and monotonically increases to $\infty$.
As $r$ increases from $2$ to $\infty$, $b_2$ monotonically increases from $-\infty$ to a local maximum $\bs_2$ at $r=r_2$ and monotonically decreases from there to $-\infty$.

\begin{figure*}[t]
\centering
\includegraphics[width=16cm]{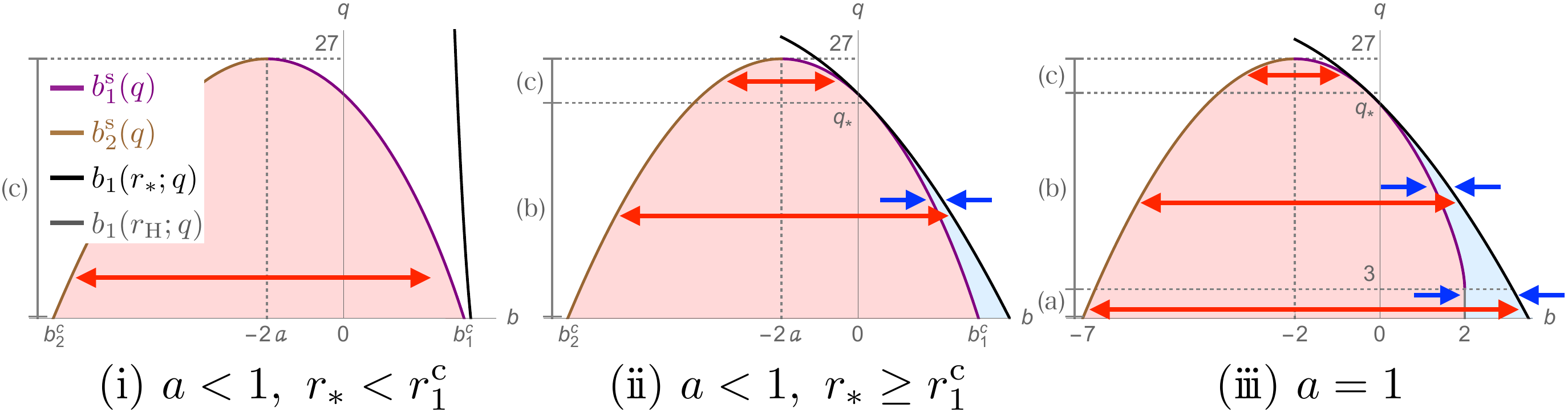}
\caption{
Typical parameter region in the $b$-$q$ plane satisfying the necessary conditions for photon escape from $r=r_*$ to infinity.
The purple, brown, and black solid curves denote $b=\bs_1(q)$, $b=\bs_2(q)$, and $b=b_1(r_*;q)$, respectively.
The gray solid segment denotes $b=b_1(\rh;q)$ and appears only for $a=1$.
The blue region represents the parameter region where photons emitted both radially inward and outward satisfy the necessary conditions for escape.
The red region represents the parameter region where photons emitted only radially outward satisfy the necessary conditions for escape.
The red and blue arrows are the same as those in Fig.~\ref{fig:b3cases}.
(i) $a=0.9$ and $r_*=\rh+10^{-2}$. 
(\ii) $a=0.9$ and $r_*=2.5$.
(\iii) $a=1$ and $r_*=2.5$.
}
\label{fig:bq_plane}
\end{figure*}

In the end, the necessary conditions for a photon to escape to infinity are given as follows. 
In Case~(a), if emitted radially outward (i.e., $\s_r=+$), a photon with $\bs_2<b\leq b_1(r_*;q)$ satisfies the necessary condition for escape to infinity [see the band between the red dashed lines in Fig.~\ref{fig:b3cases}(a)]. 
Even if emitted radially inward (i.e., $\s_r=-$), a photon with $2<b<b_1(r_*;q)$ satisfies this condition [see the band between the blue dashed lines in Fig.~\ref{fig:b3cases}(a)]. 
In Case~(b), if $\s_r=+$, a photon with $\bs_2<b\leq b_1(r_*;q)$ satisfies the necessary condition for escape to infinity [see the band between the red dashed lines in Fig.~\ref{fig:b3cases}(b)], while if $\s_r=-$, a photon with $\bs_1<b<b_1(r_*;q)$ satisfies this condition [see the band between the blue dashed lines in Fig.~\ref{fig:b3cases}(b)].
In Case~(c), a photon with $\bs_2<b<\bs_1$ satisfies the necessary condition for escape to infinity only if $\s_r=+$ [see the band between the red dashed lines in Fig.~\ref{fig:b3cases}(c)].
These are summarized in Table~\ref{table:necessary}.
It is useful to visualize the necessary conditions for photon escape in the $b$-$q$ plane. 
Figure~\ref{fig:bq_plane} shows the parameter regions necessary for photon escape.
The purple and brown curves denote $b=\bs_1(q)$ and $b=\bs_2(q)$, respectively. The black solid curve denotes $b=b_1(r_*; q)$. 
The gray segment with $b=2$ and $q\in [0, 3]$ denotes $b=b_1(\rh; q)$, which appears only for $a=1$. 
The red region shows the parameter region 
where photons emitted radially outward satisfy the necessary conditions for escape. 
The blue region shows the parameter region 
where photons emitted both radially outward and inward 
satisfy the necessary conditions for escape. 
Figures~\ref{fig:bq_plane}(i)--\ref{fig:bq_plane}(\iii) correspond to Tables~\ref{table:necessary}(i)--\ref{table:necessary}(\iii), respectively. 

%%%%%%%%%%%%%%%%%%%%%%%%%%%%%%%%%%%%%%%%%%%%%
\begin{table*}[t]
\centering
\caption{
Necessary conditions for a photon to escape to infinity.
(i) $a<1$ and $r_*<r_1^{\mathrm{c}}$. (\ii) $a<1$ and $r_* \geq r_1^{\mathrm{c}}$. (\iii) $a=1$.
}
\vspace{-5mm}
\begin{align*}
\begin{array}{cl}
(\text{i}) & 
\begin{tabular}{lccc}
\hline\hline
Case & $q$ & $b$ $(\sigma_r=+)$ & $b$ $(\sigma_r=-)$ \\ \hline
(c): $\rh<r_* \leq r_1$~ & ~$0\leq q \leq27$~ & ~$\bs_2<b<\bs_1$~ & ~not applicable \\
\hline\hline
\end{tabular} \\ & \\[-2mm]
(\text{\ii}) &
\begin{tabular}{lccc}
\hline\hline
Case & $q$ & $b$ $(\sigma_r=+)$ & $b$ $(\sigma_r=-)$ \\ \hline
(b): $\rh \leq r_1 < r_*$~ & $0\leq q<q_*$ & ~$\bs_2<b \leq b_1(r_*;q)$~ & ~$\bs_1<b<b_1(r_*;q)$ \\
(c): $\rh<r_* \leq r_1$ & ~$q_* \leq q \leq27$~ & $\bs_2<b<\bs_1$ & not applicable \\
\hline\hline
\end{tabular} \\ & \\[-2mm]
(\text{\iii}) &
\begin{tabular}{lccc}
\hline\hline
Case & $q$ & $b$ $(\sigma_r=+)$ & $b$ $(\sigma_r=-)$ \\ \hline
(a): $r_1<\rh<r_*$ & $0\leq q<3$ & $\bs_2<b \leq b_1(r_*;q)$ & $2<b<b_1(r_*;q)$ \\
(b): $\rh \leq r_1 < r_*$~ & $3\leq q<q_*$ & ~$\bs_2<b \leq b_1(r_*;q)$~ & ~$\bs_1<b<b_1(r_*;q)$ \\
(c): $\rh<r_* \leq r_1$ & ~$q_* \leq q \leq27$~ & $\bs_2<b<\bs_1$ & not applicable \\
\hline\hline
\end{tabular}
\end{array}
\end{align*}
\label{table:necessary}
\end{table*}
%%%%%%%%%%%%%%%%%%%%%%%%%%%%%%%%%%%%%%%%%%%%%
%%%%%%%%%%%%%%%%%%%%%%%%%%%%%%%%%%%%%%%%%%%%%
%%%%%%%%%%%%%%%%%%%%%%%%%%%%%%%%%%%%%%%%%%%%%

Now, let us further restrict the above necessary conditions for photon escape by the condition 
of an allowed parameter range,
\begin{align}
\Theta(\theta_*)\geq0.
\label{eq:Th*}
\end{align}
The common region of these conditions provides the necessary and sufficient parameter region in which a photon can escape to infinity. 
We call it the escapable region. 
It can be visualized in the $b$-$q$ plane, which will be a main tool to identify the escapable regions in the following sections. 
An example of the escapable region is seen in Fig.~\ref{fig:B_bq_plane}. 
The green curve denotes $\Theta(\theta_*)=0$, and the other curves and colored regions are defined in the same way as in Fig.~\ref{fig:bq_plane}. 
We can find the escapable region (i.e., the common regions colored by red and blue with $q\geq0$), which correspond to the regions in Fig.~\ref{fig:bq_plane} restricted further by the condition~\eqref{eq:Th*}.

%%%%%
\subsection{$q< 0$}
\label{subsec:q<0}
\begin{figure}[t]
\centering
\includegraphics[width=8.5cm]{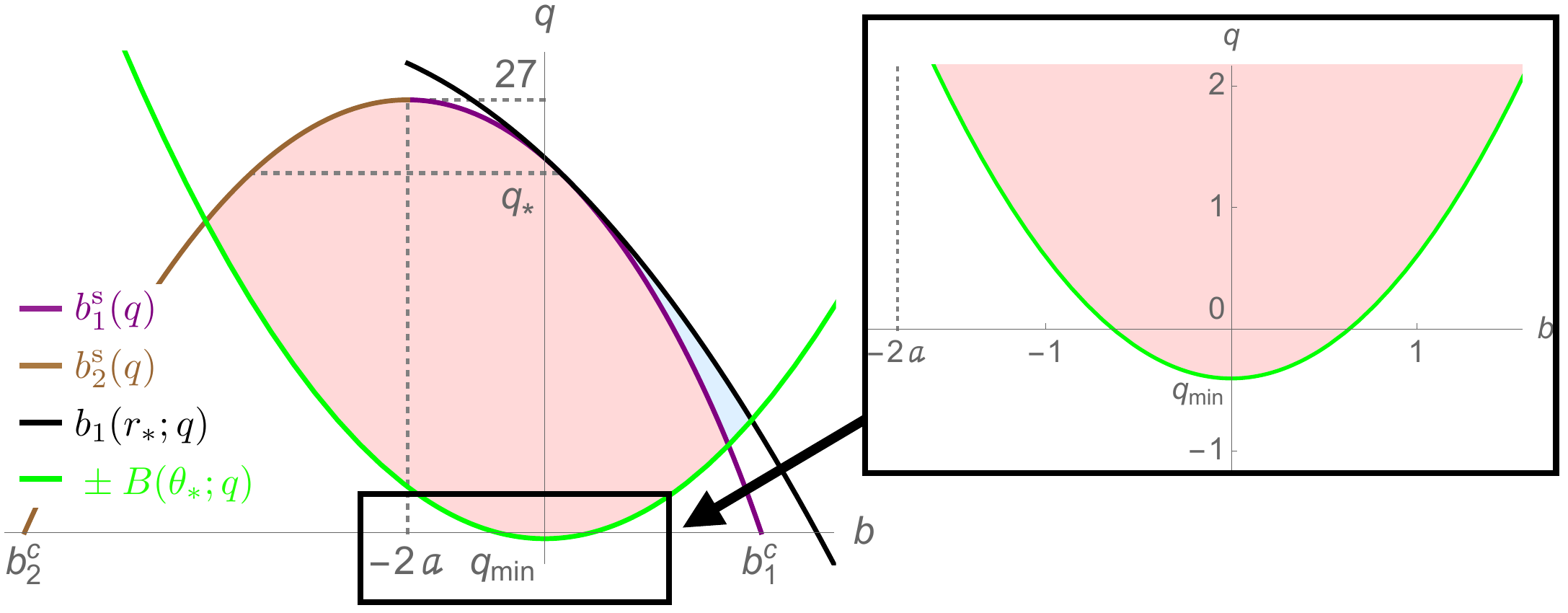}
\caption{
Typical shape of the escapable region.
The purple, brown, and black solid curves denote $b=\bs_1(q)$, $b=\bs_2(q)$, and $b=b_1(r_*;q)$, respectively.
The green solid curve denotes $\Theta(\theta_*)=0$, or equivalently, $b=\pm B(\theta_*;q)$.
The blue region represents the parameter region where photons emitted both radially inward and outward can escape to infinity.
The red region represents the parameter region where photons emitted only radially outward can escape to infinity.
}
\label{fig:B_bq_plane}
\end{figure}

%%%%%
We identify the escapable region for $q<0$. 
The negative $q$ together with 
Eq.~\eqref{q_ineq_min} at the emission point leads to
\begin{align}
  \cos^2\theta_* \left(\frac{b^2}{\sin^2\theta_*}-a^2\right)\leq q <0.
\label{eq:negq}
\end{align}
This implies that $|b|\leq a$ for $q<0$. 
The minimum value of $q$ is given at $b=0$ as 
\begin{align}
q_{\mathrm{min}}\=-a^2\cos^2\theta_*.
\end{align}
Unlike $q\geq0$, the spherical photon orbits are not relevant to photon escape because they do not exist for $q<0$. 
Therefore, we only focus on $R(r)\geq 0$, or, equivalently, 
\begin{align}
q\leq \frac{r}{\Delta}\left[\:\!
r^3+(a^2-b^2)r+2(a-b)^2
\:\!\right].
\label{eq:posiR}
\end{align}
The right-hand side is positive for all $|b|\leq a$ and $r>\rh$. 
Hence, the allowed region~\eqref{eq:posiR} contains the entire parameter region~\eqref{eq:negq}. 
This implies that any radial turning point no longer appears for $q<0$. 
Finally, we conclude that photons with negative $q$ can escape to infinity if they are emitted outwardly~(i.e., $\sigma_r=+$) and take the range \eqref{eq:negq}. 
Figure~\ref{fig:B_bq_plane} shows an example of the escapable region (see the red region of $q<0$).

%%%%%%%%%%%%%%%%%%%%%%%%%%%%%%%%%%%%%%%%%%%%%
%%%%%%%%%%%%%%%%%%%%%%%%%%%%%%%%%%%%%%%%%%%%%
\section{Critical values in the classification of photon escape}
\label{sec:criticalvalues}
%%%%%%%%%%%%%%%%%%%%%%%%%%%%%%%%%%%%%%%%%%%%%
%%%%%%%%%%%%%%%%%%%%%%%%%%%%%%%%%%%%%%%%%%%%%
%%%%%%%%%%%%%%%%%%%%%%%%%%%%%%%%%%%%%%%%%%%%%
%%%%%%%%%%%%%%%%%%%%%%%%%%%%%%%%%%%%%%%%%%%%%
\begin{figure*}[t]
\centering
\includegraphics[width=16cm]{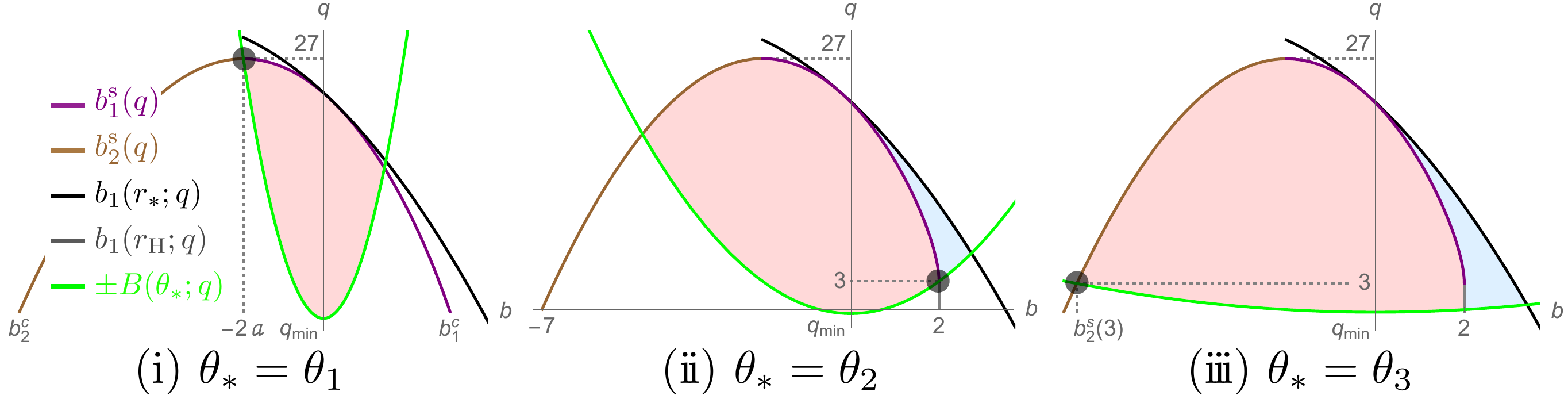}
\caption{
Escapable regions for $\theta_*$ being the critical angles (i) $\theta_*=\theta_1$, (\ii) $\theta_*=\theta_2$, and (\iii) $\theta_*=\theta_3$. The three black dots denote $(b, q)=(-2a,27)$, $(2,3)$, and $(\bs_2(3),3)$.
The meanings of the red and blue regions are the same as in Fig.~\ref{fig:B_bq_plane}.
}
\label{fig:3theta}
\end{figure*}
%%%%%%%%%%%%%%%%%%%%%%%%%%%%%%%%%%%%%%%%%%%%%

In the previous section we defined the escapable region and developed a procedure to visualize it in the $b$-$q$ plane. 
Our goal is to identify the escapable regions by indicating the allowed ranges of $b$ and $q$ explicitly.
For this purpose, in Sec.~\ref{subsec:criticalangles} we introduce three critical polar angles, at which the aspect of photon escape qualitatively changes.
Furthermore, we need to obtain the parameter values of the characteristic positions that form the boundaries of the escapable region.
In Sec.~\ref{subsec:q} we specify and name these critical parameter values.

%%%%%%%%%%%%%%%%%%%%%%%%%%%%%%%%%%%%%%%%%%%%%
\subsection{Critical angles}
\label{subsec:criticalangles}
%%%%%%%%%%%%%%%%%%%%%%%%%%%%%%%%%%%%%%%%%%%%%

We introduce critical polar angles $\theta_1$, $\theta_2$, and $\theta_3$ from three special intersections of $b=\bs_i(q)$ and $b=\pm B(\theta_*; q)$, around which the classification of the parameter values of escapable photons varies qualitatively. 
The first special point is $(b,q)=(-2a, 27)$, where $\bs_1$ and $\bs_2$ coincide with each other at $r_1=r_2=3$. 
We define $\theta_1$ by $\theta_*$ at the intersection of $b=\bs_i$ and $b=-B$, 
i.e., $-B(\theta_1;27)=-2a$ [see the black dot in Fig.~\ref{fig:3theta}(i)]. 
Then, $\theta_1$ is given by
\begin{align}
\theta_1(a)=
\arccos\left[\sqrt{\frac{3}{2a^2}\left\{\sqrt{\left(3+a^2\right)\left(27+a^2\right)} -9-a^2 \right\}}\right].
\label{def:theta_1}
\end{align}
Note that $\theta_1$ depends only on $a$ and monotonically increases with $a$ in the range 
\begin{align}
\theta_1(0)=0<\theta_1(a)\leq
\arccos\left(6\sqrt{7}-15\right)=\theta_1(1),
\end{align}
where $\arccos(6\sqrt{7}-15) \simeq 20.7^\circ$.
When $\theta_*<\theta_1$, $\bs_2<-B$ holds in the range $0\leq q \leq27$.
This implies that the minimum value of $b$ in the escapable region is always $-B$ [see Fig.~\ref{fig:3theta}(i)].

The second special point is $(b,q)=(2,3)$ for $a=1$, where $r_1=\rh$ and $\bs_1=b_1(\rh;q)=2$.
Note that $q=3$ is a special value because Cases~(a) and (b) are switched there. 
We define $\theta_2$ by $\theta_*$ at the intersection of $b=\bs_1$ and $b=B$, i.e., $B(\theta_2;3)=2$~[see the black dot in Fig.~\ref{fig:3theta}(\mbox{\ii})].
Then, $\theta_2$ is given by 
\begin{align}
\theta_2 = \arccos\left[\sqrt{2\sqrt{3}-3}\right] \simeq 47.1^\circ.
\end{align}
When $\theta_*<\theta_2$, $B<b_1(\rh;q)=2$ holds in the range $0\leq q\leq3$.
This implies that for $q\leq3$, the maximum value of $b$ in the escapable region is always $B$ [see Fig.~\ref{fig:3theta}(\mbox{\ii})].

The third special point is $(b,q)=(\bs_2(3), 3)$ for $a=1$.\\ 
We define $\theta_3$ by $\theta_*$ at the intersection of $b=-B$ and \mbox{$b=\bs_2$}, i.e., $-B(\theta_3;3)=\bs_2(3)$~[see the black dot in Fig.~\ref{fig:3theta}(\iii)].
Then, $\theta_3$ is given by
\begin{align}
\theta_3&=\arccos \Biggl[\frac{1}{\sqrt{2}}\biggl\{ \sqrt{\left(\bs_2(3)\right)^4+4\left(\bs_2(3)\right)^2+16}
\nonumber\\ &\hspace{25mm}
-2-\left(\bs_2(3)\right)^2 \biggr\}^{1/2}\Biggr]
\simeq 75.4^\circ,
\end{align}
where $\bs_2(3) \simeq -6.71$ and $r_2(3) \simeq 3.95$.
When $\theta_*<\theta_3$, $\bs_2<-B$ holds in the range $0\leq q \leq3$.
This implies that for $q \leq3$, the minimum value of $b$ in the escapable region is always $-B$ [see Fig.~\ref{fig:3theta}(\iii)].

%%%%%%%%%%%%%%%%%%%%%%%%%%%%%%%%%%%%%%%%%%%%%
\subsection{Critical values of $q$}
\label{subsec:q}
%%%%%%%%%%%%%%%%%%%%%%%%%%%%%%%%%%%%%%%%%%%%%
%%%%%%%%%%%%%%%%%%%%%%%%%%%%%%%%%%%%%%%%%%%%%
\begin{table*}[t]
\centering
\caption{
Definitions of the critical values of $q$ and conditions that appear. 
}
\begin{tabular}{lll}
\hline\hline
Critical value $q$ \hspace{5mm} & Definition & Necessary conditions to appear \\ \hline
$q_1(\theta_*)$ & $b_1(\rh;q)=B(\theta_*;q)$ & only for $a=1$ and $\theta_* \geq \theta_2$ \\
$q_2(r_*,\theta_*)$ & $b_1(r_*;q)=B(\theta_*;q)$ & always \\
$q_3(\theta_*)$ & $\bs_1(q)=B(\theta_*;q)$ & other than $a=1$ and $\theta_* \geq \theta_2$ \\
$q_4(\theta_*)$ & $\bs_2(q)=-B(\theta_*;q)$ & only for $\theta_* \geq \theta_1$ \\
$q_5(\theta_*)$ & $\bs_1(q)=-B(\theta_*;q)$ & only for $\theta_*<\theta_1$ \\
$q_6(r_*,\theta_*)$ \hspace{5mm} & $b_1(r_*;q)=-B(\theta_*;q)$ \hspace{5mm} & 
only for $\theta_*<\theta_1$ \\
\hline\hline
\end{tabular}
\label{table:critical_q}
\end{table*}
%%%%%%%%%%%%%%%%%%%%%%%%%%%%%%%%%%%%%%%%%%%%%
%%%%%%%%%%%%%%%%%%%%%%%%%%%%%%%%%%%%%%%%%%%%%
\begin{figure*}[t]
\centering
\includegraphics[width=12cm]{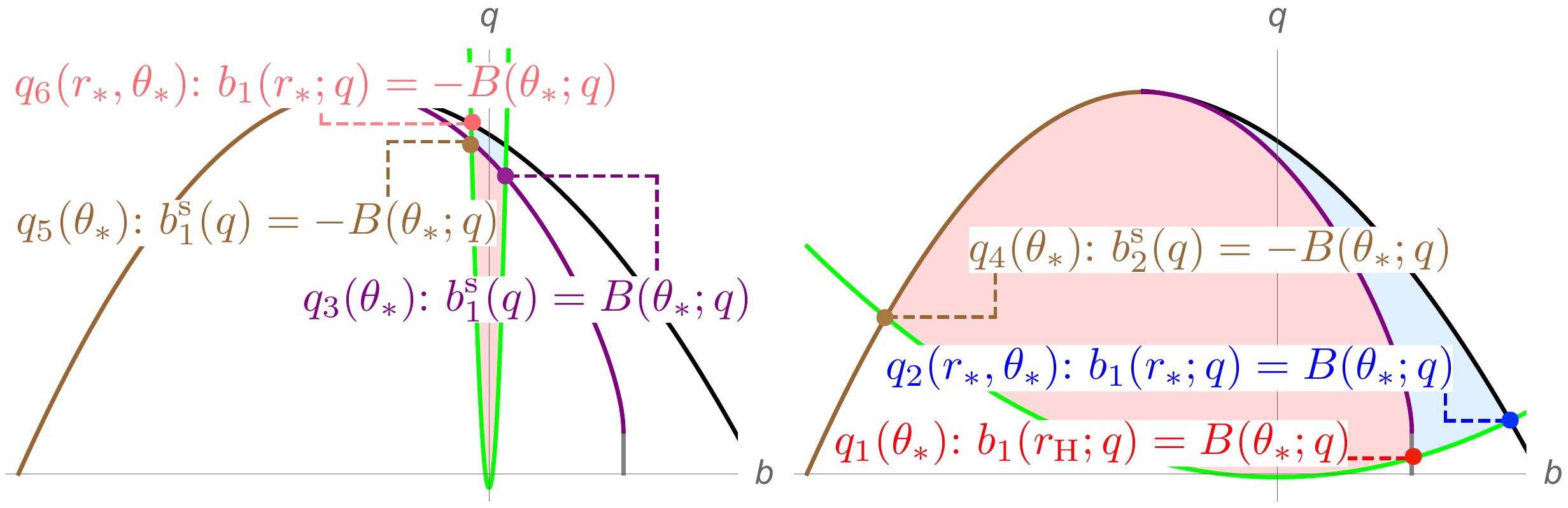}
\caption{
Six critical values of $q$.
}
\label{fig:6q}
\end{figure*}
%%%%%%%%%%%%%%%%%%%%%%%%%%%%%%%%%%%%%%%%%%%%%

We introduce six critical values of $q$ from the special intersections of $b=b_1(\rh;q)$, $b=b_1(r_*;q)$, $b=\bs_i(q)$, and $b=\pm B(\theta_*;q)$, at which the classification of the parameter ranges for photon escape varies qualitatively.

We define $q_1$ as the value of $q$ at the intersection of $b=b_1(\rh;q)$ and $b=B(\theta_*;q)$ for $a=1$ [see the red dot in Fig.~\ref{fig:6q}], 
\begin{align}
q_1(\theta_*)=\frac{3+\cos^2\theta_*}{\tan^2\theta_*},
\end{align}
which only appears for $\theta_* \in [\theta_2, \pi/2)$ and monotonically decreases with $\theta_*$ in the range $q_1(\pi/2)=0<q_1(\theta_*)\leq 3=q_1(\theta_2)$.
When $q<q_1$, then $B<b_1(\rh;q)$ holds.
This implies that for $q<q_1$, the maximum value of $b$ in the escapable region is always $B$.

%%%%%%%%%%%%%%%%%%%%%%%%%%%%%%%%%%%%%%%%%%%%%
\begin{figure*}[t]
\centering
\includegraphics[width=12cm]{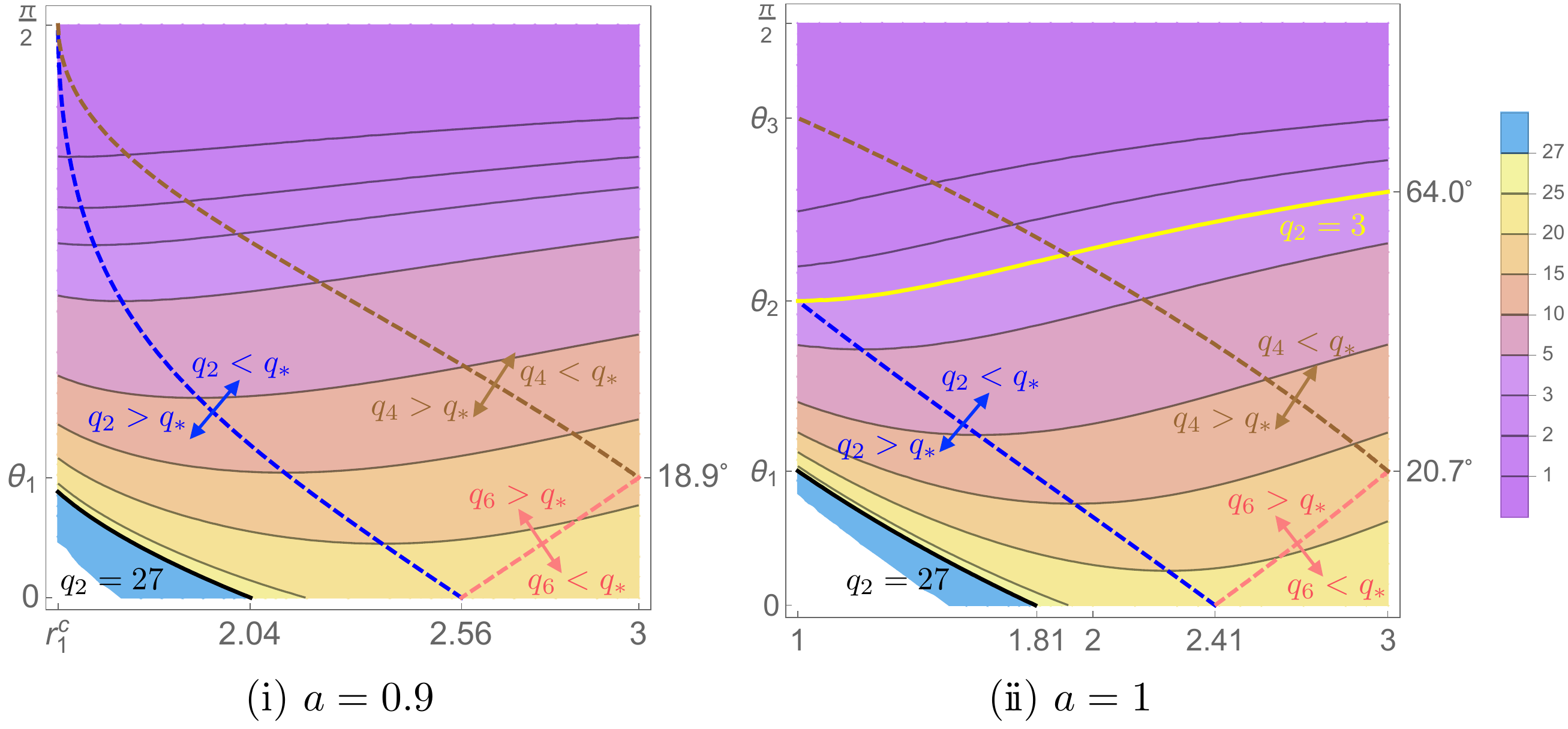}
\caption{
Value of $q_2(r_*,\theta_*)$ in the $r_*$-$\theta_*$ parameter space. The black and yellow solid curves denote $q_2=27$ and $q_2=3$, respectively. 
The blue dashed curve gives the minimum value of $q_2$ for fixed $\theta_*$ and satisfies $q_*=q_2(r_*,\theta_*)=q_3(\theta_*)$.
The brown and pink dashed curves satisfy $q_*=q_4(\theta_*)$ and $q_*=q_5(\theta_*)=q_6(r_*,\theta_*)$, respectively. 
(i) $a=0.9$. (\ii) $a=1$.
}
\label{fig:q2_ContourPlot}
\end{figure*}
%%%%%%%%%%%%%%%%%%%%%%%%%%%%%%%%%%%%%%%%%%%%%
%%%%%%%%%%%%%%%%%%%%%%%%%%%%%%%%%%%%%%%%%%%%%
\begin{figure*}[t]
\centering
\includegraphics[width=12cm]{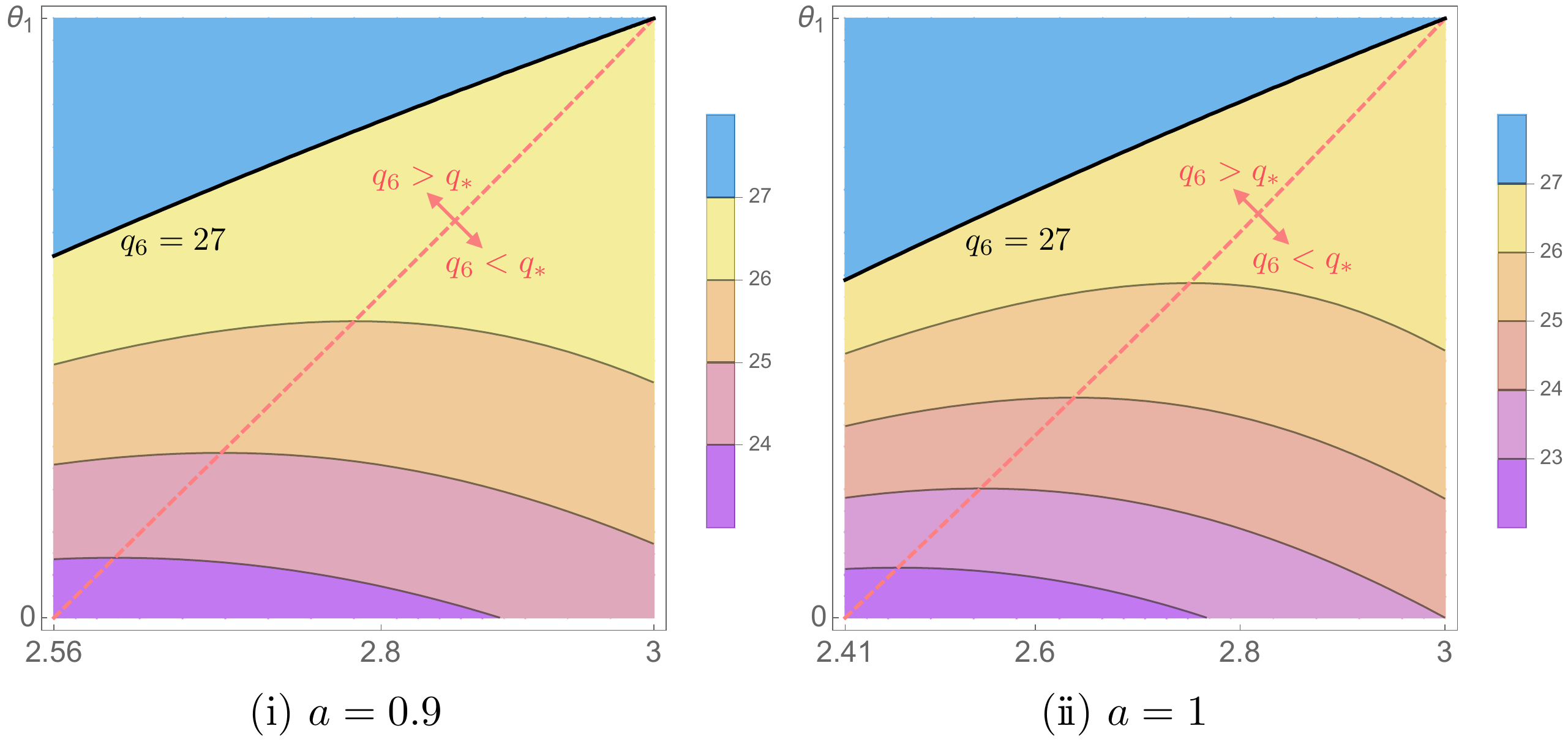}
\caption{
Value of $q_6(r_*,\theta_*)$ in the $r_*$-$\theta_*$ parameter space.
The black solid curve denotes $q_2=27$. 
The pink dashed curve gives the minimum value of $q_6$ for fixed $\theta_*$ and satisfies $q_*=q_6(r_*,\theta_*)=q_5(\theta_*)$. 
(i) $a=0.9$. (\ii) $a=1$.
}
\label{fig:q6_ContourPlot}
\end{figure*}
%%%%%%%%%%%%%%%%%%%%%%%%%%%%%%%%%%%%%%%%%%%%%

We define $q_2$ as the value of $q$ at the intersection of $b=b_1(r_*;q)$ and $b=B(\theta_*;q)$ [see the blue dot in Fig.~\ref{fig:6q}],
\begin{align}
q_2(r_*,\theta_*)=\cos^2\theta_*
\left[\left(\frac{\S_*\sqrt{\D_*}-2ar_*\sin\theta_*}{\D_*-a^2\sin^2\theta_*}\right)^2-a^2\right],
\end{align}
where $\S_*\=r^2_*+a^2\cos^2\theta_*$ and $\D_*\=r^2_*-2r_*+a^2$.
Note that $q_2$ depends on both $r_*$ and $\theta_*$, and if $r_*$ is fixed it monotonically decreases with $\theta_*$, but if $\theta_*$ is fixed it is not always monotonic with $r_*$.
Figure~\ref{fig:q2_ContourPlot} shows the value of $q_2$ in the $r_*$-$\theta_*$ parameter space.
When $q_2$ exists in the range $q_2 \leq 27$ and $q$ is in the range $q<q_2$, then $B<b_1(r_*;q)$ holds.
This implies that for $q<q_2$, the maximum value of $b$ in the escapable region is always $B$.

We define $q_3$ as the value of $q$ at the intersection of $b=\bs_1(q)$ and $b=B(\theta_*;q)$ [see the purple dot in Fig.~\ref{fig:6q}].
When $a<1$, $q_3$ always appears, while when $a=1$ it only appears for $\theta_* \in (0,\theta_2)$.
Note that $q_3(\theta_*)$ monotonically decreases with $\theta_*$ in the range $q_3(\pi/2)=0<q_3(\theta_*)<q_3(0)$ for $a<1$, while $q_3(\theta_2)=3<q_3(\theta_*)<11+8\sqrt{2}=q_3(0)$ for $a=1$.
When $q<q_3$, then $B<\bs_1$ holds.
This implies that for $q<q_3$, the maximum value of $b$ in the escapable region is always $B$.

We define $q_4$ as the value of $q$ at the intersection of $b=\bs_2(q)$ and $b=-B(\theta_*;q)$ [see the brown dot in Fig.~\ref{fig:6q}], which only appears for $\theta_*\in [\theta_1,\pi/2)$.
Note that $q_4(\theta_*)$ monotonically decreases with $\theta_*$ in the range $q_4(\pi/2)=0<q_4(\theta_*) \leq 27=q_4(\theta_1)$.
When $q<q_4$, then $\bs_2<-B$ holds.
This implies that for $q<q_4$, the minimum value of $b$ in the escapable region is always $-B$.

We define $q_5$ as the value of $q$ at the intersection of $b=\bs_1(q)$ and $b=-B(\theta_*;q)$ [see the brown dot in Fig.~\ref{fig:6q}], which only appears for $\theta_*\in(0, \theta_1)$.
Note that $q_5(\theta_*)$ monotonically increases with $\theta_*$ in the range 
$q_5(0)<q_5(\theta_*) < 27=q_5(\theta_1)$.
When $q_*<q_5$ or $r_*<r_1^{\mathrm{c}}$, the maximum value of $q$ in the escapable region is $q_5$, i.e., there is no escapable region for $q\geq q_5$.

We define $q_6$ as the value of $q$ at the intersection of $b=b_1(r_*;q)$ and $b=-B(\theta_*;q)$ [see the pink dot in Fig.~\ref{fig:6q}],
\begin{align}
q_6(r_*,\theta_*)=\cos^2\theta_*
\left[\left(\frac{\S_*\sqrt{\D_*}+2ar_*\sin\theta_*}{\D_*-a^2\sin^2\theta_*}\right)^2-a^2\right].
\end{align}
Note that $q_6$ depends on both $r_*$ and $\theta_*$, and if $r_*$ is fixed it monotonically increases with $\theta_*$, but if $\theta_*$ is fixed it is not monotonic with $r_*$.
Figure~\ref{fig:q6_ContourPlot} shows the value of $q_6$ in the $r_*$-$\theta_*$ parameter space.
When $q_6\leq q_*$, the maximum value of $q$ in the escapable region is $q_6$, i.e., there is no escapable region for $q\geq q_6$.

These six critical $q$ values are summarized in Table~\ref{table:critical_q}.
It is worth noting that the critical values always satisfy the following inequalities:
\begin{align}
\label{eq:ineq356}
    &q_3<q_5\leq q_6,
    \\
    \label{eq:ineq326}
    &q_3\leq q_2<q_6,
    \\
    \label{eq:ineq324}
    &q_3\leq q_2<q_4,
    \\
    \label{eq:ineq124}
    &q_1<q_2<q_4.
\end{align}
In the following sections, we will perform a complete classification of photon escape.

%%%%%%%%%%%%%%%%%%%%%%%%%%%%%%%%%%%%%%%%%%%%%
%%%%%%%%%%%%%%%%%%%%%%%%%%%%%%%%%%%%%%%%%%%%%
\section{Escapable region in an extremal Kerr black hole}
\label{sec:classification_ext}
%%%%%%%%%%%%%%%%%%%%%%%%%%%%%%%%%%%%%%%%%%%%%

In this section we make a complete classification of photon escape in an extremal Kerr black hole.
In this case, we define four classes according to $\theta_*$:
Class~I, $0<\theta_*<\theta_1$; Class~\2, $\theta_1\leq\theta_*<\theta_2$; Class~\3, $\theta_2\leq\theta_*<\theta_3$; Class~\4, $\theta_3\leq\theta_*<\pi/2$ (see Table~\ref{table:classes_a=1}).

%%%%%%%%%%%%%%%%%%%%%%%%%%%%%%%%%%%%%%%%%%%%%
\begin{table}[t]
\centering
\caption{
($a=1$) Definition of each Class and the critical values of $q$ in an extremal Kerr black hole.
}
\begin{tabular}{lll}
\hline\hline
Class & Range of $\theta_*$ & Critical values of $q$\\ \hline
Class~I & $0<\theta_*<\theta_1$ & $q_2$, $q_3$, $q_5$, and $q_6$ \\
Class~\2 & $\theta_1\leq\theta_*<\theta_2$ & $q_2$, $q_3$, and $q_4$ \\
Class~\3 & $\theta_2\leq\theta_*<\theta_3$ & $q_1$, $q_2$, and $q_4$ \\
Class~\4 \hspace{1mm} & $\theta_3\leq\theta_*<\pi/2$ \hspace{1mm} & $q_1$, $q_2$, and $q_4$ \\
\hline\hline
\end{tabular}
\label{table:classes_a=1}
\end{table}
%%%%%%%%%%%%%%%%%%%%%%%%%%%%%%%%%%%%%%%%%%%%%

%%%%%%%%%%%%%%%%%%%%%%%%%%%%%%%%%%%%%%%%%%%%%
\subsection{Class~I: $0<\theta_*<\theta_1$ and $a=1$}
\label{subsec:ClassI}
%%%%%%%%%%%%%%%%%%%%%%%%%%%%%%%%%%%%%%%%%%%%%

In Class~I, there are five characteristic $q$'s:
$q_*=q_{\mathrm{SPO}}(r_*)$, $q_2$, $q_3$, $q_5$, and $q_6$.
Here $q_3$ monotonically decreases with $\theta_*$, and $q_5$ monotonically increases with $\theta_*$ in the ranges
\begin{align}
q_3(\theta_1)<q_3(\theta_*)<q_3(0) \text{ and }
q_5(0)<q_5(\theta_*)<q_5(\theta_1),
\end{align}
respectively, where $q_3(\theta_1) \simeq 12.6$, $q_3(0)=q_5(0)=11+8\sqrt{2}$, and $q_5(\theta_1)=27$.
If $r_*$ is fixed, $q_2$ monotonically decreases with $\theta_*$.
If $\theta_*$ is fixed, as $r_*$ increases from $1$ to $3$, $q_2$ monotonically decreases from $q_2(1,\theta_*)$ to a local minimum $q_2=q_*=q_3$ at $r_*=r_1(q_3)$ and monotonically increases from there to $q_2(3,\theta_*)$
[see the regions $1<r_*<3$ and $0<\theta_*<\theta_1$ in Fig.~\ref{fig:q2_ContourPlot}(\ii)].

%%%%%%%%%%%%%%%%%%%%%%%%%%%%%%%%%%%%%%%%%%%%%
\begin{table}[t]
\centering
\caption{(Class~I, $a=1$)
Escapable region $(b,q)$ for an extremal Kerr black hole with $0<\theta_*<\theta_1$. 
(Class~I-2, $a<1$)
Escapable region $(b,q)$ for a subextremal Kerr black hole with $r_* \geq r_1^{\mathrm{c}}$ and $0<\theta_*<\theta_1$.
}
\begin{tabular}{lccc}
\hline\hline
Case & $q$ & $b$ $(\sigma_r=+)$ & $b$ $(\sigma_r=-)$ \\ \hline
(i), (\ii) & $q_{\mathrm{min}}\leq q<q_3$ & $-B\leq b\leq B$ & not applicable \\
& $q_3\leq q<q_5$ & $-B\leq b< \bs_1$ & not applicable \\
    & $q_5\leq q\leq27$ & not applicable & not applicable \\\hline
(\iii) & $q_{\mathrm{min}}\leq q< q_3$ & $-B\leq b\leq B$ & not applicable \\
    &$q_3\leq q< q_2$ & $-B\leq b\leq B$ & $\bs_1<b\leq B$ \\
    &$q_2\leq q<q_*$ & $-B\leq b \leq b_1(r_*;q)$ & $\bs_1<b<b_1(r_*;q)$ \\
    &$q_*\leq q<q_5$ & $-B\leq b < \bs_1$ & not applicable \\
    &$q_5\leq q\leq27$ & not applicable & not applicable \\\hline
(\iv) & $q_{\mathrm{min}}\leq q< q_3$ & $-B\leq b\leq B$ & not applicable \\
    &$q_3\leq q< q_2$ & $-B\leq b\leq B$ & $\bs_1<b\leq B$ \\
    &$q_2\leq q<q_6$ & $-B\leq b \leq b_1(r_*;q)$ & $\bs_1<b<b_1(r_*;q)$ \\
    &$q_6\leq q\leq27$ & not applicable & not applicable \\\hline
(v) & $q_{\mathrm{min}}\leq q< q_3$ & $-B\leq b\leq B$ & not applicable \\
    &$q_3\leq q< q_2$ & $-B\leq b\leq B$ & $\bs_1<b\leq B$ \\
    &$q_2\leq q<q_5$ & $-B\leq b \leq b_1(r_*;q)$ & $\bs_1<b<b_1(r_*;q)$ \\
    &$q_5\leq q<q_6$ & $-B\leq b \leq b_1(r_*;q)$ & $-B\leq b < b_1(r_*;q)$ \\
    &$q_6\leq q\leq27$ & not applicable & not applicable \\\hline
(\vi) & $q_{\mathrm{min}}\leq q< q_3$ & $-B\leq b\leq B$ & not applicable \\
    &$q_3\leq q< q_5$ & $-B\leq b\leq B$ & $\bs_1<b\leq B$ \\
    &$q_5\leq q<q_6$ & $-B\leq b \leq b_1(r_*;q)$ & $-B\leq b < b_1(r_*;q)$ \\
    &$q_6\leq q\leq27$ & not applicable & not applicable \\\hline
(\vii) & $q_{\mathrm{min}}\leq q< q_3$ & $-B\leq b\leq B$ & not applicable \\
    &$q_3\leq q< q_5$ & $-B\leq b\leq B$ & $\bs_1<b\leq B$ \\
    &$q_5\leq q<q_2$ & $-B\leq b \leq B$ & $-B\leq b\leq B$ \\
    &$q_2\leq q<q_6$ & $-B\leq b \leq b_1(r_*;q)$ & $-B\leq b < b_1(r_*;q)$ \\
    &$q_6\leq q\leq27$ & not applicable & not applicable \\
\hline\hline
\end{tabular}
\label{table:ClassI}
\end{table}
%%%%%%%%%%%%%%%%%%%%%%%%%%%%%%%%%%%%%%%%%%%%%
%%%%%%%%%%%%%%%%%%%%%%%%%%%%%%%%%%%%%%%%%%%%%
\begin{figure}[t]
\centering
\includegraphics[width=8.5cm]{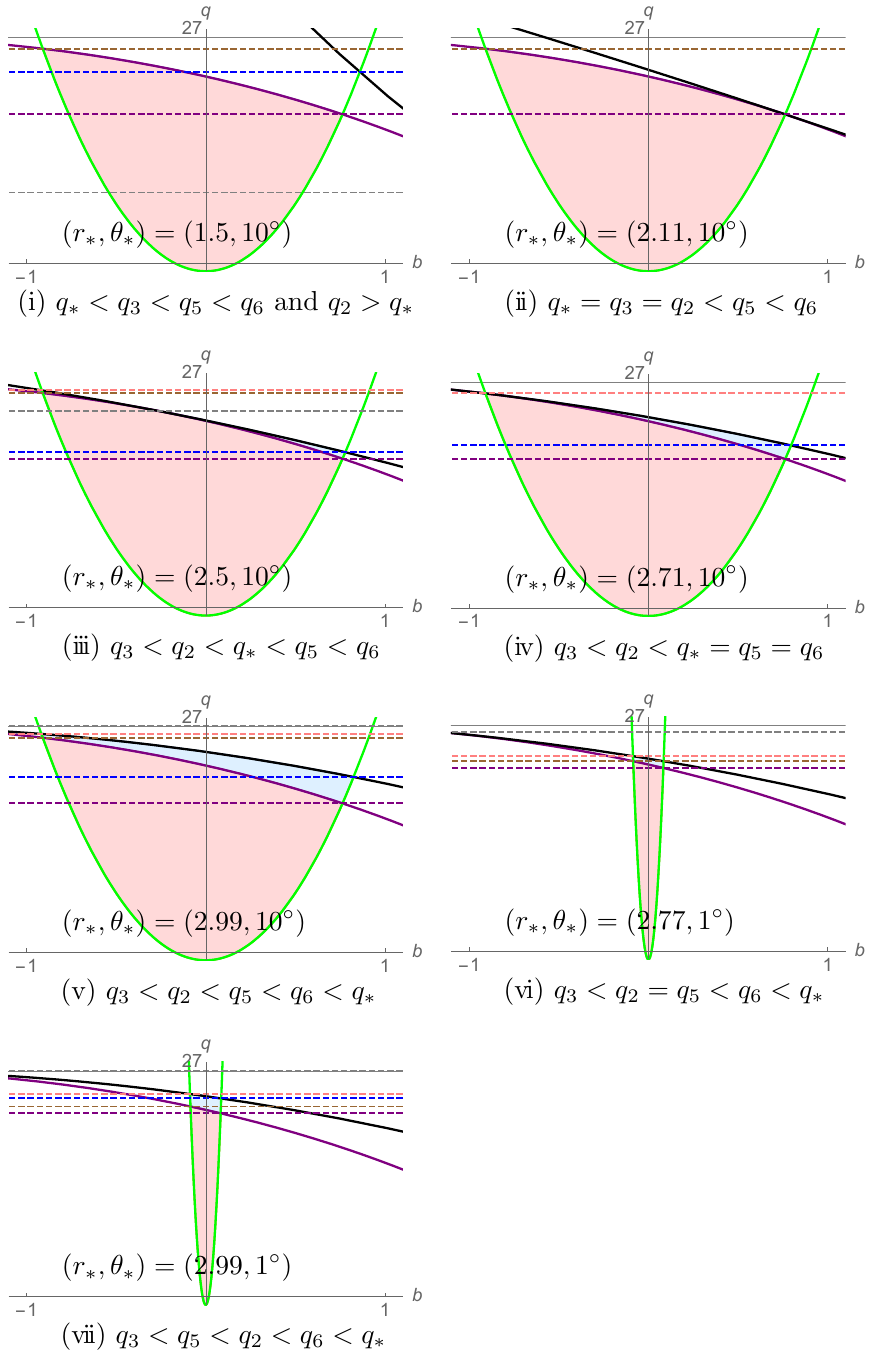}
\caption{(Class~I, $a=1$)
Typical shape of the escapable region for an extremal Kerr black hole with $0<\theta_*<\theta_1$. The purple, black, and green curves denote $b=\bs_1(q)$, $b_1(r_*;q)$, and $\pm B(\theta_*;q)$, respectively.
The gray, blue, purple, brown, and pink dashed lines denote $q=q_*$, $q_2$, $q_3$, $q_5$, and $q_6$, respectively.
}
\label{fig:ClassI_bq_plane}
\end{figure}
%%%%%%%%%%%%%%%%%%%%%%%%%%%%%%%%%%%%%%%%%%%%%

These critical values satisfy the inequalities~\eqref{eq:ineq356} and \eqref{eq:ineq326}. 
Therefore, there exist seven cases according to the relative values of $q_*$ to $q_2$, $q_3$, $q_5$, and $q_6$:
\begin{align}
\begin{array}{cl}
\text{(i)} & q_*<q_3<q_5<q_6 \text{ and } q_2>q_*, \\
\text{(\ii)} & q_*=q_3=q_2<q_5<q_6, \\
\text{(\iii)} & q_3<q_2<q_*<q_5<q_6, \\
\text{(\iv)} & q_3<q_2<q_*=q_5=q_6, \\
\text{(v)} & q_3<q_2<q_5<q_6<q_*, \\
\text{(\vi)} & q_3<q_2=q_5<q_6<q_*, \\
\text{(\vii)} & q_3<q_5<q_2<q_6<q_*.
\end{array}
\label{relative_q_cases_ClassI}
\end{align}
Note that Cases~(\vi) and (\vii) appear only when \mbox{$\theta_*<2.54^\circ$}.

It is worth noting that when $q_2>q_*$, the intersection of $b=b_1(r_*;q)$ and $b=B(\theta_*;q)$ does not contribute to specifying the escapable region
[see Fig.~\ref{fig:ClassI_bq_plane}(i)].
On the other hand, when $q_2 \leq q_*$, 
the intersection of $b=b_1(r_*;q)$ and $b=B(\theta_*;q)$ is a special point, where the shape of the escapable region changes
[see, e.g., Fig.~\ref{fig:ClassI_bq_plane}(\iii)].
Therefore, we need to consider $q_2$ for specifying the escapable region only when $q_2 \leq q_*$.
In particular, in Case~(i) $q_2$ can take three ranges: $q_2<q_5$, $q_2=q_5$, and $q_2>q_5$. 
However, since $q_2>q_*$ in all three ranges, we do not distinguish them.

For the same reason as for $q_2$, the relative values of $q_*$ and $q_6$ determine whether $q_6$ contributes to specifying the escapable region.
When $q_6>q_*$, the intersection of $b=b_1(r_*;q)$ and $b=-B(\theta_*;q)$ is not included in the escapable region~[see Fig.~\ref{fig:ClassI_bq_plane}(\iii)].
On the other hand, when $q_6 \leq q_*$, the intersection of $b=b_1(r_*;q)$ and $b=-B(\theta_*;q)$ is a special point, where the shape of the escapable region changes~[see, e.g., Fig.~\ref{fig:ClassI_bq_plane}(v)].
Therefore, we need to consider $q_6$ only when $q_6\leq q_*$.

The escapable regions in the above cases are summarized in Table~\ref{table:ClassI} and Fig.~\ref{fig:ClassI_bq_plane}.

%%%%%%%%%%%%%%%%%%%%%%%%%%%%%%%%%%%%%%%%%%%%%
\subsection{Class~\2: $\theta_1\leq\theta_*<\theta_2$ and $a=1$}
%%%%%%%%%%%%%%%%%%%%%%%%%%%%%%%%%%%%%%%%%%%%%

In Class~\2, there are four characteristic $q$'s: $q_*$, $q_2$, $q_3$, and $q_4$. 
Here $q_3$ and $q_4$ monotonically decrease with $\theta_*$ in the ranges
\begin{align}
q_3(\theta_2)<q_3(\theta_*) \leq q_3(\theta_1) \text{ and }
q_4(\theta_2)<q_3(\theta_*) \leq q_4(\theta_1),
\end{align}
respectively, where $q_3(\theta_2)=3$, $q_3(\theta_1) \simeq 12.6$, $q_4(\theta_2) \simeq 18.9$, and $q_4(\theta_1)=27$.
The behavior of $q_2$ is the same as in Class~I, and the value of $q_2$ is in the range $3<q_2<27$
[see the regions of $1<r_*<3$ and $\theta_1\leq \theta_*<\theta_2$ in Fig.~\ref{fig:q2_ContourPlot}(\mbox{\ii})].

%%%%%%%%%%%%%%%%%%%%%%%%%%%%%%%%%%%%%%%%%%%%%
\begin{table}[t]
\centering
\caption{(Class~\2, $a=1$)
Escapable region $(b,q)$ for an extremal Kerr black hole with $\theta_1 \leq \theta_*<\theta_2$.
(Class~\2-2, $a<1$)
Escapable region $(b,q)$ for a subextremal Kerr black hole with $r_* \geq r_1^{\mathrm{c}}$ and $\theta_1 \leq \theta_*<\pi/2$.}
\begin{tabular}{lccc}
\hline\hline
Case & $q$ & $b$ $(\sigma_r=+)$ & $b$ $(\sigma_r=-)$ \\ \hline
(i), (\ii) & $q_{\mathrm{min}}\leq q<q_3$ & $-B\leq b\leq B$ & not applicable \\
    & $q_3\leq q<q_4$ & $-B\leq b< \bs_1$ & not applicable \\
    & $q_4\leq q\leq27$ & $\bs_2<b<\bs_1$ & not applicable \\\hline
(\iii) & $q_{\mathrm{min}}\leq q< q_3$ & $-B\leq b\leq B$ & not applicable \\
    &$q_3\leq q< q_2$ & $-B\leq b\leq B$ &  $\bs_1<b\leq B$ \\
    &$q_2\leq q<q_*$ & $-B\leq b \leq b_1(r_*;q)$ & $\bs_1<b<b_1(r_*;q)$ \\
    &$q_*\leq q<q_4$ & $-B\leq b < \bs_1$ & not applicable \\
    &$q_4\leq q\leq27$ & $\bs_2<b<\bs_1$ & not applicable \\\hline
(\iv) & $q_{\mathrm{min}}\leq q< q_3$ & $-B\leq b\leq B$ & not applicable \\
    &$q_3\leq q< q_2$ & $-B\leq b\leq B$ & $\bs_1<b\leq B$ \\
    &$q_2\leq q<q_4$ & $-B\leq b \leq b_1(r_*;q)$ & $\bs_1<b<b_1(r_*;q)$ \\
    &$q_4\leq q\leq27$ & $\bs_2<b<\bs_1$ & not applicable \\\hline
(v) & $q_{\mathrm{min}}\leq q< q_3$ & $-B\leq b\leq B$ & not applicable \\
    &$q_3\leq q< q_2$ & $-B\leq b\leq B$ & $\bs_1<b\leq B$ \\
    &$q_2\leq q<q_4$ & $-B\leq b \leq b_1(r_*;q)$ & $\bs_1<b<b_1(r_*;q)$ \\
    &$q_4\leq q<q_*$ & $\bs_2< b \leq b_1(r_*;q)$ & $\bs_1<b<b_1(r_*;q)$ \\
    &$q_*\leq q\leq27$ & $\bs_2<b<\bs_1$ & not applicable \\
\hline\hline
\end{tabular}
\label{table:ClassII}
\end{table}
%%%%%%%%%%%%%%%%%%%%%%%%%%%%%%%%%%%%%%%%%%%%%
%%%%%%%%%%%%%%%%%%%%%%%%%%%%%%%%%%%%%%%%%%%%%
\begin{figure}[t]
\centering
\includegraphics[width=8.5cm]{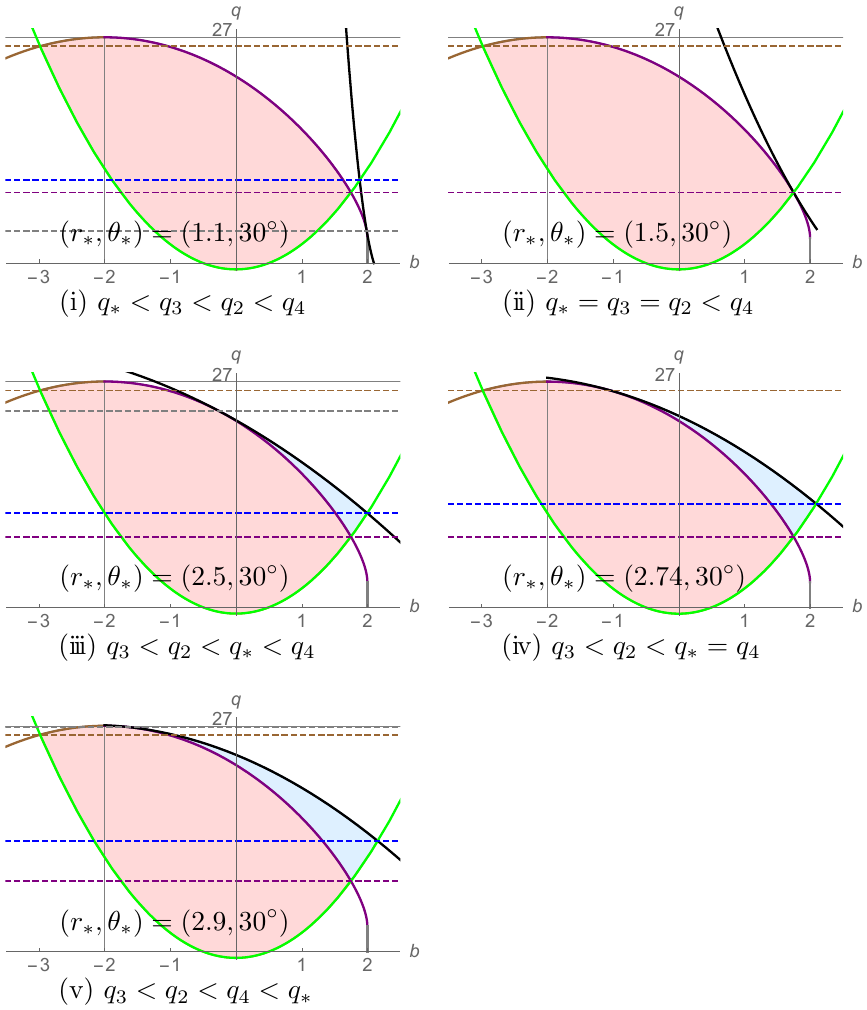}
\caption{(Class~\2, $a=1$)
Typical shape of the escapable region for an extremal Kerr black hole with $\theta_1\leq\theta_*<\theta_2$. The purple, brown, black, gray, and green curves denote $b=\bs_1(q)$, $\bs_2(q)$, $b_1(r_*;q)$, 
$b_1(\rh;q)$ and $\pm B(\theta_*;q)$, respectively.
The gray, blue, purple, and brown dashed lines denote $q=q_*$, $q_2$, $q_3$, and $q_4$, respectively.
}
\label{fig:ClassII_bq_plane}
\end{figure}
%%%%%%%%%%%%%%%%%%%%%%%%%%%%%%%%%%%%%%%%%%%%%

These critical values satisfy the inequalities~\eqref{eq:ineq324}.
Therefore, there exist five cases according to the relative values of $q_*$ to $q_2$, $q_3$, and $q_4$:
\begin{align}
\begin{array}{cl}
\text{(i)} & q_*<q_3<q_2<q_4, \\
\text{(\ii)} & q_*=q_3=q_2<q_4, \\
\text{(\iii)} & q_3<q_2<q_*<q_4, \\
\text{(\iv)} & q_3<q_2<q_*=q_4, \\
\text{(v)} & q_3<q_2<q_4<q_*. 
\end{array}
\label{relative_q_cases_ClassII}
\end{align}
The escapable regions in the above cases are summarized in Table~\ref{table:ClassII} and Fig.~\ref{fig:ClassII_bq_plane}.

%%%%%%%%%%%%%%%%%%%%%%%%%%%%%%%%%%%%%%%%%%%%%
\subsection{Class~\3: $\theta_2\leq\theta_*<\theta_3$ and $a=1$}
%%%%%%%%%%%%%%%%%%%%%%%%%%%%%%%%%%%%%%%%%%%%%

In Class~\3, there are four characteristic $q$'s: $q_*$, $q_1$, $q_2$, and $q_4$. Here $q_1$ and $q_4$ monotonically decrease with $\theta_*$ in the ranges
\begin{align}
q_1(\theta_3)<q_1(\theta_*) \leq q_1(\theta_2) \text{ and }
q_4(\theta_3)<q_4(\theta_*) \leq q_4(\theta_2),
\end{align}
respectively, where $q_1(\theta_3) \simeq 0.208$, $q_1(\theta_2)=3$, $q_4(\theta_3)=3$, and $q_4(\theta_2) \simeq 18.9$.
The critical value $q_2$ monotonically decreases with $\theta_*$ for fixed $r_*$ and monotonically increases with $r_*$ for fixed $\theta_*$, and hence we have 
\begin{align}
&q_2(1, \theta_*)<q_2<q_2(3,\theta_*),\\
&q_2(r_*, \theta_3)<q_2<q_2(r_*,\theta_2),
\end{align}
where the minimum value is $q_2(1, \theta_3)\simeq 0.208$ and the maximum value is $q_2(3,\theta_2)\simeq 7.71$ [see the regions of $1<r_*<3$ and $\theta_2\leq \theta_*<\theta_3$ in Fig.~\ref{fig:q2_ContourPlot}(\ii)].

%%%%%%%%%%%%%%%%%%%%%%%%%%%%%%%%%%%%%%%%%%%%%
\begin{table}[t]
\centering
\caption{(Class~\3, $a=1$)
Escapable region $(b,q)$ for an extremal Kerr black hole with $\theta_2 \leq \theta_*<\theta_3$.
}
\begin{tabular}{lccc}
\hline\hline
Case & $q$ & $b$ $(\sigma_r=+)$ & $b$ $(\sigma_r=-)$ \\ \hline
(i) & $q_{\mathrm{min}}\leq q<q_1$ & $-B\leq b\leq B$ & not applicable\\
    & $q_1\leq q\leq 3$ & $-B\leq b\leq B$ & $2<b\leq B$ \\
    & $3\leq q<q_2$ & $-B\leq b\leq B$ & $\bs_1<b\leq B$ \\
    & $q_2\leq q<q_*$ & $-B\leq b\leq b_1(r_*;q)$ & $\bs_1<b<b_1(r_*;q)$ \\
    & $q_*\leq q<q_4$ & $-B\leq b<\bs_1$ & not applicable \\
    & $q_4\leq q\leq 27$ & $\bs_2<b<\bs_1$ & not applicable \\ \hline
(\ii) & $q_{\mathrm{min}}\leq q<q_1$ & $-B\leq b\leq B$ & not applicable \\
    & $q_1\leq q\leq 3$ & $-B\leq b\leq B$ & $2<b\leq B$ \\
    & $3\leq q<q_2$ & $-B\leq b\leq B$ & $\bs_1<b\leq B$ \\
    & $q_2\leq q<q_4$ & $-B\leq b\leq b_1(r_*;q)$ & $\bs_1<b<b_1(r_*;q)$ \\
    & $q_4\leq q\leq 27$ & $\bs_2<b<\bs_1$ & not applicable \\ \hline
(\iii) & $q_{\mathrm{min}}\leq q<q_1$ & $-B\leq b\leq B$ & not applicable \\
    & $q_1\leq q\leq 3$ & $-B\leq b\leq B$ & $2<b \leq B$ \\
    & $3\leq q<q_2$ & $-B\leq b\leq B$ & $\bs_1<b\leq B$\\
    & $q_2\leq q<q_4$ & $-B\leq b\leq b_1(r_*;q)$ & $\bs_1<b<b_1(r_*; q)$\\
    & $q_4\leq q<q_*$ & $\bs_2<b \leq b_1(r_*;q)$ & $\bs_1<b<b_1(r_*;q)$ \\
    & $q_*\leq q\leq 27$ & $\bs_2<b<\bs_1$ & not applicable \\ \hline
(\iv) & $q_{\mathrm{min}}\leq q<q_1$ & $-B\leq b\leq B$ & not applicable \\
    & $q_1\leq q<3$ & $-B\leq b\leq B$ & $2<b \leq B$ \\
    & $3\leq q<q_*$ & $-B\leq b\leq b_1(r_*;q)$ & $\bs_1<b<b_1(r_*;q)$ \\
    & $q_*\leq q<q_4$ & $-B\leq b<\bs_1$ & not applicable \\
    & $q_4\leq q\leq 27$ & $\bs_2<b<\bs_1$ & not applicable \\ \hline
(v) & $q_{\mathrm{min}}\leq q<q_1$ & $-B\leq b\leq B$ & not applicable \\
    & $q_1\leq q<3$ & $-B\leq b\leq B$ & $2<b \leq B$ \\
    & $3\leq q<q_4$ & $-B\leq b\leq b_1(r_*;q)$ & $\bs_1<b<b_1(r_*;q)$ \\
    & $q_4\leq q\leq 27$ & $\bs_2<b<\bs_1$ & not applicable \\ \hline
(\vi) & $q_{\mathrm{min}}\leq q<q_1$ & $-B\leq b\leq B$ & not applicable \\
    & $q_1\leq q<3$ & $-B\leq b\leq B$ & $2<b \leq B$ \\
    & $3\leq q<q_4$ & $-B\leq b\leq b_1(r_*;q)$ & $\bs_1<b<b_1(r_*;q)$ \\
    & $q_4\leq q< q_*$ & $\bs_2<b\leq b_1(r_*;q)$ & $\bs_1<b<b_1(r_*;q)$ \\
    & $q_*\leq q\leq 27$ & $\bs_2<b<\bs_1$ & not applicable \\ \hline
(\vii) & $q_{\mathrm{min}}\leq q<q_1$ & $-B\leq b\leq B$ & not applicable \\
    & $q_1\leq q<q_2$ & $-B\leq b\leq B$ & $2<b \leq B$ \\
    & $q_2\leq q<3$ & $-B\leq b\leq b_1(r_*;q)$ & $2<b<b_1(r_*;q)$ \\
    & $3\leq q<q_*$ & $-B\leq b\leq b_1(r_*;q)$ & $\bs_1<b<b_1(r_*;q)$ \\
    & $q_*\leq q<q_4$ & $-B\leq b<\bs_1$ & not applicable \\
    & $q_4\leq q\leq 27$ & $\bs_2<b<\bs_1$ & not applicable \\ \hline
(\viii) & $q_{\mathrm{min}}\leq q<q_1$ & $-B\leq b\leq B$ & not applicable \\
    & $q_1\leq q<q_2$ & $-B\leq b\leq B$ & $2<b \leq B$ \\
    & $q_2\leq q<3$ & $-B\leq b\leq b_1(r_*;q)$ & $2<b<b_1(r_*;q)$ \\
    & $3\leq q<q_4$ & $-B\leq b\leq b_1(r_*;q)$ & $\bs_1<b<b_1(r_*;q)$ \\
    & $q_4\leq q\leq 27$ & $\bs_2<b<\bs_1$ & not applicable \\ \hline
(\ix) & $q_{\mathrm{min}}\leq q<q_1$ & $-B\leq b\leq B$ & not applicable \\
    & $q_1\leq q<q_2$ & $-B\leq b\leq B$ & $2<b \leq B$ \\
    & $q_2\leq q<3$ & $-B\leq b\leq b_1(r_*;q)$ & $2<b<b_1(r_*;q)$ \\
    & $3\leq q<q_4$ & $-B\leq b\leq b_1(r_*;q)$ & $\bs_1<b<b_1(r_*;q)$ \\
    & $q_4\leq q<q_*$ & $\bs_2<b \leq b_1(r_*;q)$ & $\bs_1<b<b_1(r_*;q)$ \\
    & $q_*\leq q\leq 27$ & $\bs_2<b<\bs_1$ & not applicable \\ 
\hline\hline
\end{tabular}
\label{table:ClassIII}
\end{table}
%%%%%%%%%%%%%%%%%%%%%%%%%%%%%%%%%%%%%%%%%%%%%
%%%%%%%%%%%%%%%%%%%%%%%%%%%%%%%%%%%%%%%%%%%%%
\begin{figure}[t]
\centering
\includegraphics[width=8.5cm]{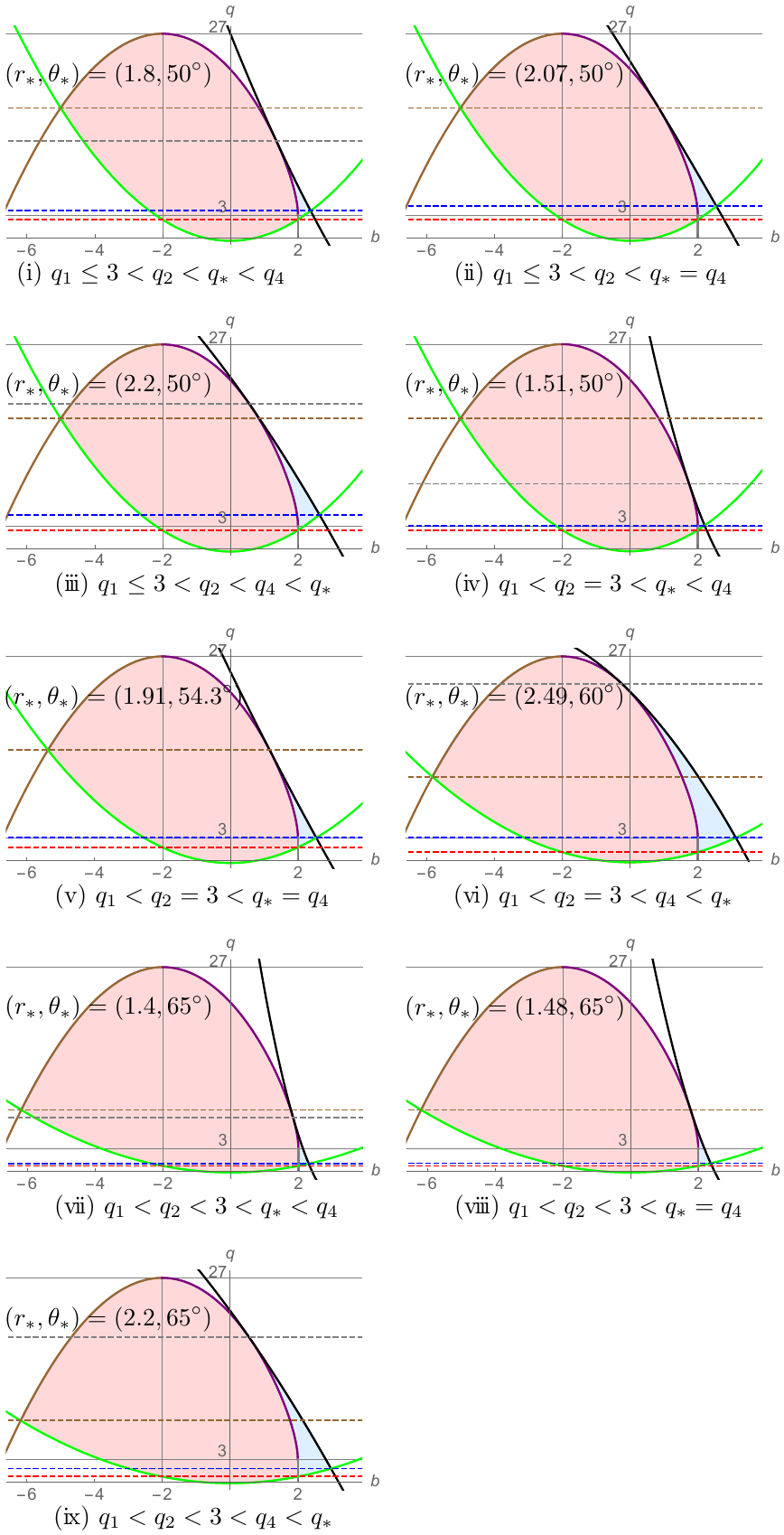}
\caption{(Class~\3, $a=1$)
Typical shape of the escapable region for an extremal Kerr black hole with $\theta_2 \leq \theta_*<\theta_3$. The purple, brown, black, gray, and green curves denote $b=\bs_1(q)$, $\bs_2(q)$, $b_1(r_*;q)$, $b_1(\rh;q)$, and $\pm B(\theta_*;q)$, respectively.
The gray, red, blue, and brown dashed lines denote $q=q_*$, $q_1$, $q_2$, and $q_4$, respectively.
}
\label{fig:ClassIII_bq_plane}
\end{figure}
%%%%%%%%%%%%%%%%%%%%%%%%%%%%%%%%%%%%%%%%%%%%%

These critical values satisfy 
\begin{align}
&q_1\leq 3<q_4,\\
&q_1<q_2<q_4, \\
&q_2<q_*,
\end{align}
where the second inequalities correspond to Eq.~\eqref{eq:ineq124}.
Therefore, there exist nine cases according to the relative values of $q_*$ to $q_1$, $q_2$, $q_4$, and $q_{\mathrm{SPO}}(\rh)=3$:
\begin{align}
\begin{array}{cl}
\text{(i)} & q_1 \leq 3<q_2<q_*<q_4, \\
\text{(\ii)} & q_1 \leq 3<q_2<q_*=q_4, \\
\text{(\iii)} & q_1 \leq 3<q_2<q_4<q_*, \\
\text{(\iv)} & q_1<q_2=3<q_*<q_4, \\
\text{(v)} & q_1<q_2=3<q_*=q_4, \\
\text{(\vi)} & q_1<q_2=3<q_4<q_*, \\
\text{(\vii)} & q_1<q_2<3<q_*<q_4, \\
\text{(\viii)} & q_1<q_2<3<q_*=q_4, \\
\text{(\ix)} & q_1<q_2<3<q_4<q_*.
\end{array}
\end{align}
The escapable regions in the above cases are summarized in Table~\ref{table:ClassIII} and Fig.~\ref{fig:ClassIII_bq_plane}.

\clearpage

%%%%%%%%%%%%%%%%%%%%%%%%%%%%%%%%%%%%%%%%%%%%%
\subsection{Class~\4: $\theta_3\leq\theta_*<\pi/2$ and $a=1$}
%%%%%%%%%%%%%%%%%%%%%%%%%%%%%%%%%%%%%%%%%%%%%

\begin{figure}[t]
\centering
\includegraphics[width=8.5cm]{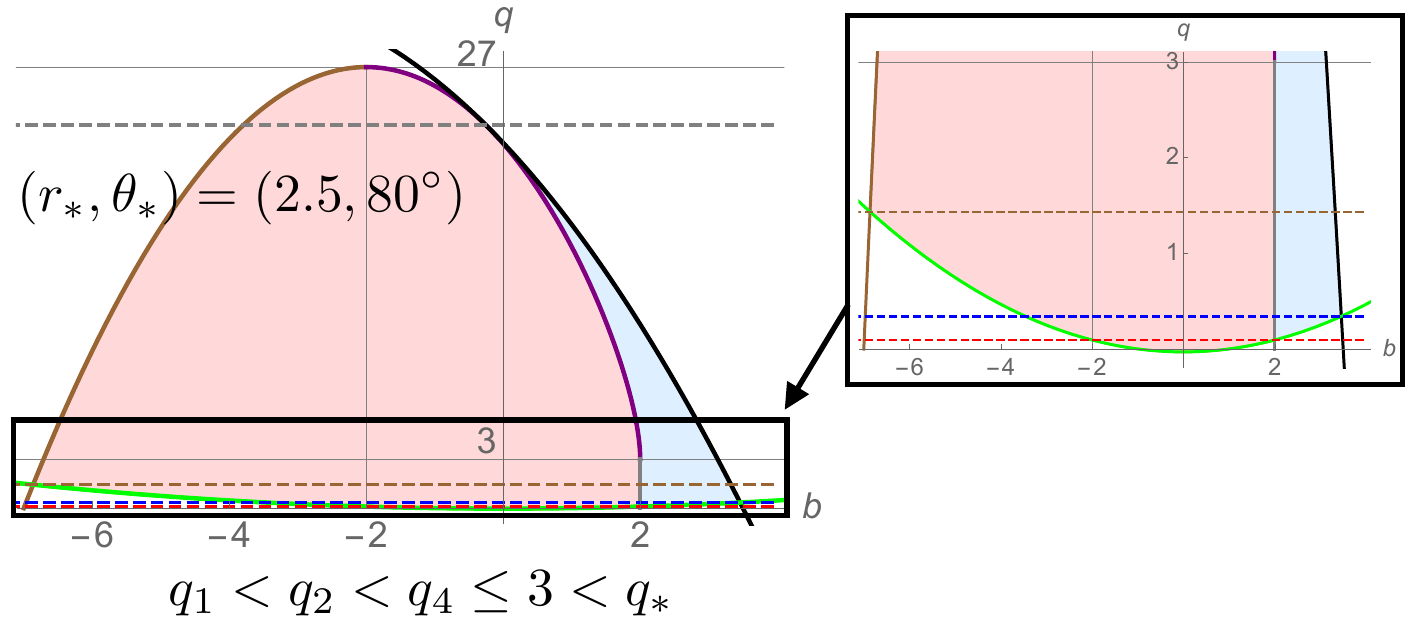}
\caption{(Class~\4, $a=1$)
Typical shape of the escapable region for an extremal Kerr black hole with $\theta_3 \leq \theta_*<\pi/2$. The purple, brown, black, gray, and green curves denote $b=\bs_1(q)$, $\bs_2(q)$, $b_1(r_*;q)$, $b_1(\rh;q)$, and $\pm B(\theta_*;q)$, respectively.
The gray, red, blue, and brown dashed lines denote $q=q_*$, $q_1$, $q_2$, and $q_4$, respectively.
}
\label{fig:ClassIV_bq_plane}
\end{figure}

In Class~\4, there are four characteristic $q$'s: $q_*$, $q_1$, $q_2$, and $q_4$.
Since these $q$'s satisfy the inequality
\begin{align}
q_1<q_2<q_4\leq3<q_*,
\end{align}
there is no case classification according to the relative values of $q$.
The escapable regions are summarized in Table~\ref{table:ClassIV} and Fig.~\ref{fig:ClassIV_bq_plane}.

%%%%%%%%%%%%%%%%%%%%%%%%%%%%%%%%%%%%%%%%%%%%%
\begin{table}[t]
\centering
\caption{(Class~\4, $a=1$)
Escapable region $(b,q)$ for an extremal Kerr black hole with $\theta_3 \leq \theta_*<\pi/2$.
}
\begin{tabular}{ccc}
\hline\hline
$q$ & $b$ $(\sigma_r=+)$ & $b$ $(\sigma_r=-)$ \\ \hline
$q_{\mathrm{min}}\leq q<q_1$ & $-B\leq b\leq B$ & not applicable \\
$q_1\leq q<q_2$ & $-B\leq b\leq B$ & $2<b \leq B$ \\
$q_2\leq q<q_4$ & $-B\leq b\leq b_1(r_*;q)$ & $2<b<b_1(r_*;q)$ \\
$q_4\leq q\leq 3$& $\bs_2<b\leq b_1(r_*;q)$ & $2<b<b_1(r_*;q)$ \\
$3\leq q<q_*$ & $\bs_2<b\leq b_1(r_*;q)$ & $\bs_1<b<b_1(r_*;q)$ \\
$q_*\leq q\leq 27$ & $\bs_2<b<\bs_1$ & not applicable \\
\hline\hline
\end{tabular}
\label{table:ClassIV}
\end{table}
%%%%%%%%%%%%%%%%%%%%%%%%%%%%%%%%%%%%%%%%%%%%%
%%%%%%%%%%%%%%%%%%%%%%%%%%%%%%%%%%%%%%%%%%%%%
%%%%%%%%%%%%%%%%%%%%%%%%%%%%%%%%%%%%%%%%%%%%%

%%%%%%%%%%%%%%%%%%%%%%%%%%%%%%%%%%%%%%%%%%%%%
%%%%%%%%%%%%%%%%%%%%%%%%%%%%%%%%%%%%%%%%%%%%%
\section{Escapable region in a subextremal Kerr black hole}
\label{sec:classification_sub}
%%%%%%%%%%%%%%%%%%%%%%%%%%%%%%%%%%%%%%%%%%%%%
%%%%%%%%%%%%%%%%%%%%%%%%%%%%%%%%%%%%%%%%%%%%%
\begin{table}[b]
\centering
\caption{
($a<1$) Definition of each Class and the critical values of $q$ in a subextremal Kerr black hole.}
\begin{tabular}{llll}
\hline\hline
Class & Range of $r_*$ \hspace{.3mm} & Range of $\theta_*$ & Critical values of $q$ \\ \hline
Class~I-1 & $r_*<r_1^{\mathrm{c}}$ & $0<\theta_*<\theta_1$ & $q_3$ and $q_5$ \\
Class~I-2 & $r_* \geq r_1^{\mathrm{c}}$ & $0<\theta_*<\theta_1$ & $q_2$, $q_3$, $q_5$, and $q_6$ \\
Class~\2-1 \hspace{.3mm} & $r_*<r_1^{\mathrm{c}}$ & $\theta_1\leq\theta_*<\pi/2$ \hspace{.3mm} & $q_3$ and $q_4$ \\
Class~\2-2 & $r_* \geq r_1^{\mathrm{c}}$ & $\theta_1\leq\theta_*<\pi/2$& $q_2$, $q_3$, and $q_4$ \\
\hline\hline
\end{tabular}
\label{table:classes_a<1}
\end{table}
%%%%%%%%%%%%%%%%%%%%%%%%%%%%%%%%%%%%%%%%%%%%%

In this section we make a complete classification of photon escape in a subextremal Kerr black hole.
We again note that when $a<1$ and $r_*<r_1^{\mathrm{c}}$, $b=b_1(r_*;q)$ does not intersect with $b=\bs_1(q)$ in the $b$-$q$ plane, and thus $q_*$ does not appear 
[see Fig.~\ref{fig:bq_plane}(i) and Table~\ref{table:necessary}(i)].
Therefore, though the class for $a=1$ is only defined by the range of $\theta_*$, the class for $a<1$ is defined by the ranges of $r_*$ and $\theta_*$ as follows:
Class~I-1, $r_*<r_1^{\mathrm{c}}$ and $0<\theta_*<\theta_1$; Class~I-2, $r_* \geq r_1^{\mathrm{c}}$ and $0<\theta_*<\theta_1$; \mbox{Class~\2-1}, $r_*<r_1^{\mathrm{c}}$ and $\theta_1\leq\theta_*<\pi/2$; Class~\2-2, $r_* \geq r_1^{\mathrm{c}}$ and $\theta_1\leq\theta_*<\pi/2$ (see Table~\ref{table:classes_a<1}).

%%%%%%%%%%%%%%%%%%%%%%%%%%%%%%%%%%%%%%%%%%%%%
\subsection{Classes~I-1 and \2-1: $\rh<r_*<r_1^{\mathrm{c}}$ and $a<1$}
\begin{figure}[t]
\centering
\includegraphics[width=8.5cm]{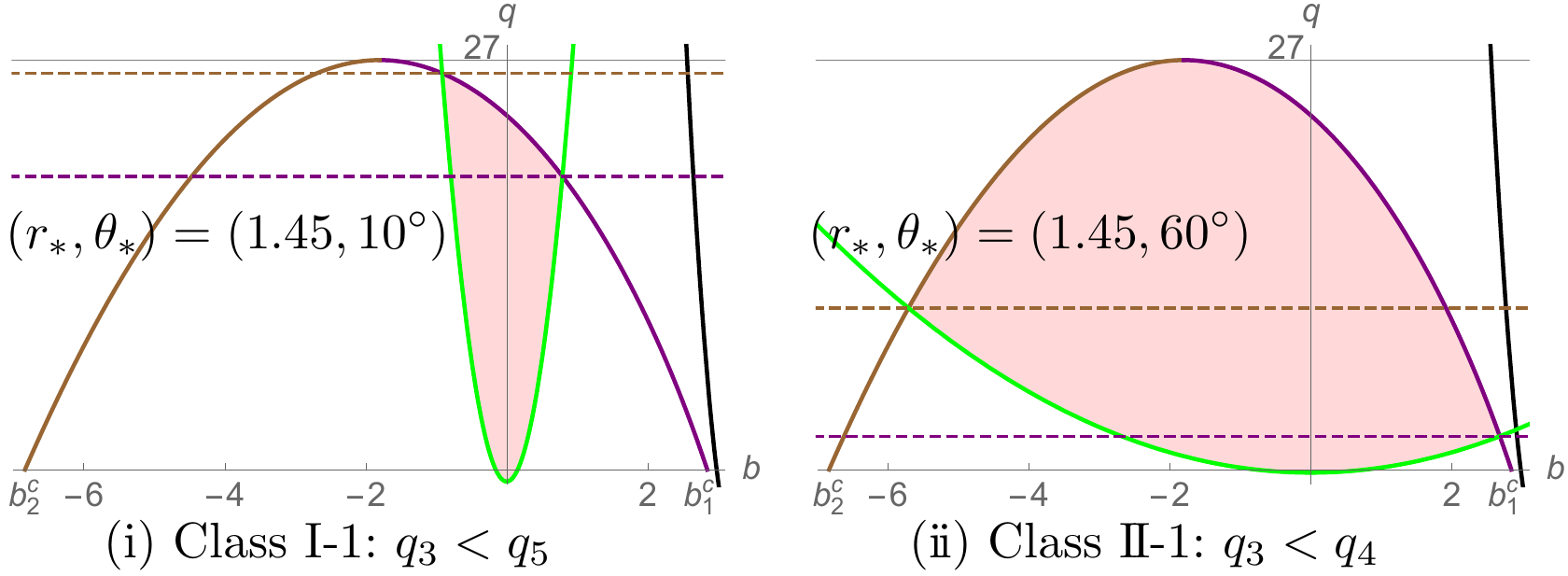}
\caption{(Class~I-1 and Class~\2-1, $a<1$)
Typical shape of the escapable region for a subextremal Kerr black hole with $r_*<r_1^{\mathrm{c}}$. The purple, brown, black, and green curves denote $b=\bs_1(q)$, $\bs_2(q)$, $b_1(r_*;q)$, and $\pm B(\theta_*;q)$, respectively. Here we set $a=0.9$, and then $\rh\simeq 1.44$ and $r_1^{\mathrm{c}}\simeq 1.56$.
(i) $\theta_*$ is in the range $0<\theta_*<\theta_1$ (Class~I-1).
The purple and brown dashed lines denote $q=q_3$ and $q_5$, respectively.
(\ii) $\theta_*$ is in the range $\theta_1 \leq \theta*<\pi/2$ (Class~\2-1).
The purple and brown dashed lines denote $q=q_3$ and $q_4$, respectively.
}
\label{fig:ClassIaIIa_bq_plane}
\end{figure}

%%%%%%%%%%%%%%%%%%%%%%%%%%%%%%%%%%%%%%%%%%%%%
In Class~I-1, there are only two characteristic $q$'s: $q_3$ and $q_5$.
Since $q_3<q_5$, there is no case classification according to the relative values of critical $q$.
Similarly, in Class~\2-1 there are only two characteristic $q$, $q_3$ and $q_4$.
Since $q_3<q_4$, there is no case classification according to the relative values of critical $q$.
The escapable regions are summarized in Table~\ref{table:ClassIaIIa_sub} and Fig.~\ref{fig:ClassIaIIa_bq_plane}.

%%%%%%%%%%%%%%%%%%%%%%%%%%%%%%%%%%%%%%%%%%%%%
\begin{table}[t]
\centering
\caption{(Class~I-1 and Class~\2-1, $a<1$)
Escapable region $(b,q)$ with $r_*<r_1^{\mathrm{c}}$ and $0<\theta_*<\theta_1$ (Class~I-1) and $r_*<r_1^{\mathrm{c}}$ and $\theta_1\leq\theta_*<\pi/2$ (Class~\2-1) for a subextremal Kerr black hole.
}
\begin{tabular}{lccc}
\hline\hline
Class & $q$ & $b$ $(\sigma_r=+)$ & $b$ $(\sigma_r=-)$ \\ \hline
I-1 & $q_{\mathrm{min}}\leq q<q_3$ & $-B\leq b\leq B$ & not applicable \\
& $q_3 \leq q<q_5$ & $-B \leq b<\bs_1$ & not applicable \\
& $q_5 \leq q \leq 27$ & not applicable & not applicable \\ \hline
\2-1 & $q_{\mathrm{min}}\leq q<q_3$ & $-B\leq b\leq B$ & not applicable \\
& $q_3 \leq q<q_4$ & $-B \leq b<\bs_1$ & not applicable  \\
& $q_4 \leq q \leq 27$ & $\bs_2<b<\bs_1$ & not applicable \\
\hline\hline
\end{tabular}
\label{table:ClassIaIIa_sub}
\end{table}
%%%%%%%%%%%%%%%%%%%%%%%%%%%%%%%%%%%%%%%%%%%%%
%%%%%%%%%%%%%%%%%%%%%%%%%%%%%%%%%%%%%%%%%%%%%
%%%%%%%%%%%%%%%%%%%%%%%%%%%%%%%%%%%%%%%%%%%%%

\newpage

%%%%%%%%%%%%%%%%%%%%%%%%%%%%%%%%%%%%%%%%%%%%%
\subsection{Class~I-2: $r_* \geq r_1^{\mathrm{c}}$, $0<\theta_*<\theta_1$ and $a<1$}
%%%%%%%%%%%%%%%%%%%%%%%%%%%%%%%%%%%%%%%%%%%%%

\begin{figure}[t]
\centering
\includegraphics[width=8.5cm]{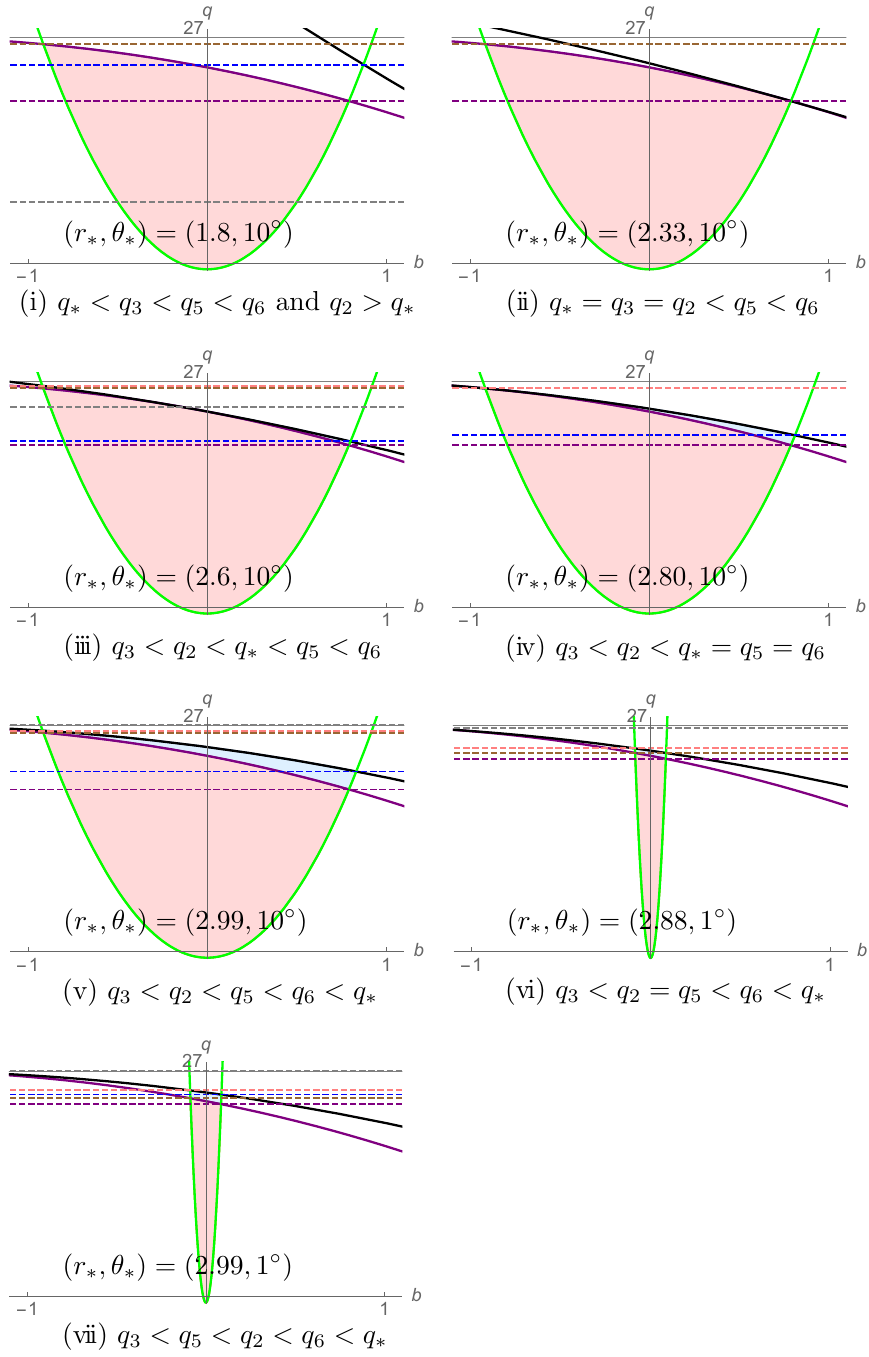}
\caption{(Class~I-2, $a<1$)
Typical shape of the escapable region for a subextremal Kerr black hole with $r_* \geq r_1^{\mathrm{c}}$ and $0<\theta_*<\theta_1$. The purple, black, and green curves denote $b=\bs_1(q)$, $b_1(r_*;q)$, and $\pm B(\theta_*;q)$, respectively.
The gray, blue, purple, brown, and pink dashed lines denote 
$q=q_*$, $q_2$, $q_3$, $q_5$, and $q_6$ respectively.
Here we set $a=0.9$, and then $r_1^{\mathrm{c}}\simeq 1.56$.
}
\label{fig:ClassIb_bq_plane}
\end{figure}

In Class~I-2, there are five characteristic $q$'s: $q_*$, $q_2$, $q_3$, $q_5$, and $q_6$.
They are classified into seven cases according to the relative values, which are the same as the cases~\eqref{relative_q_cases_ClassI} of Class~I for $a=1$.
The escapable regions are summarized in Table~\ref{table:ClassI} and Fig.~\ref{fig:ClassIb_bq_plane}.

%%%%%%%%%%%%%%%%%%%%%%%%%%%%%%%%%%%%%%%%%%%%%
%%%%%%%%%%%%%%%%%%%%%%%%%%%%%%%%%%%%%%%%%%%%%

%%%%%%%%%%%%%%%%%%%%%%%%%%%%%%%%%%%%%%%%%%%%%
\subsection{Class~\2-2: $r_* \geq r_1^{\mathrm{c}}$, $\theta_1 \leq \theta_*<\pi/2$, and $a<1$}
%%%%%%%%%%%%%%%%%%%%%%%%%%%%%%%%%%%%%%%%%%%%%

\begin{figure}[t]
\centering
\includegraphics[width=8.5cm]{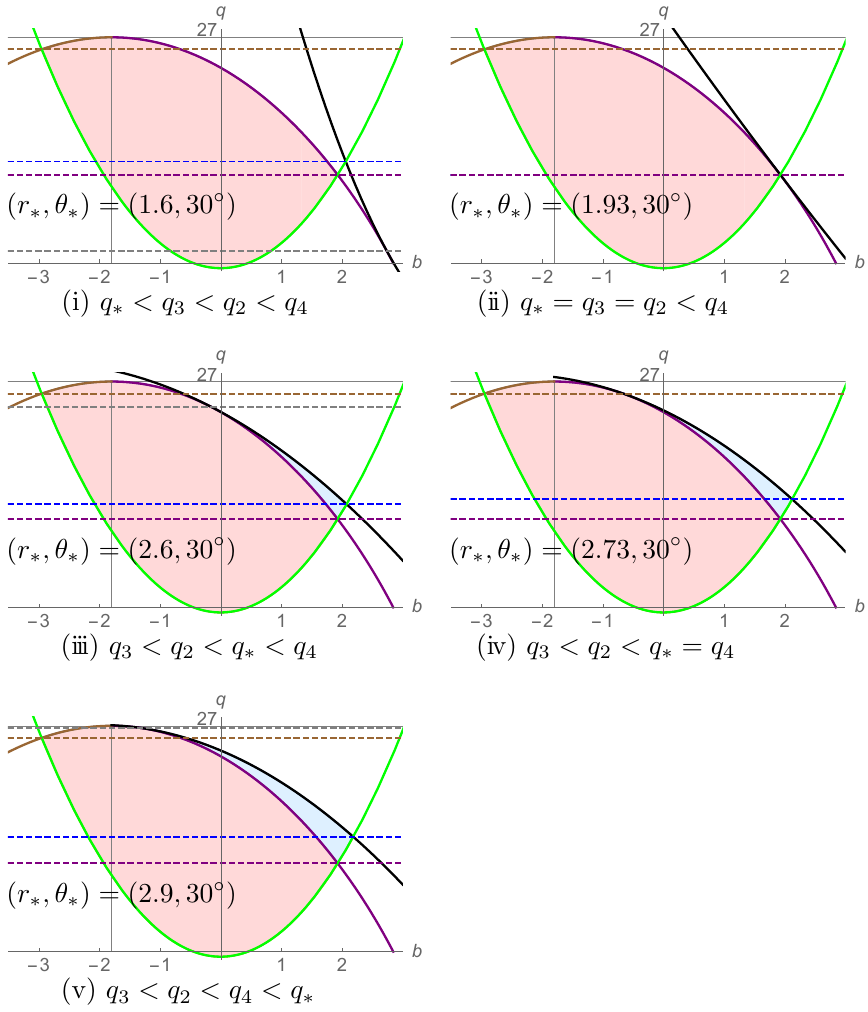}
\caption{(Class~\2-2, $a<1$)
Typical shape of the escapable region for a subextremal Kerr black hole with $r_* \geq r_1^{\mathrm{c}}$ and $\theta_1\leq\theta_*<\pi/2$. The purple, brown, black, and green curves denote $b=\bs_1(q)$, $\bs_2(q)$, $b_1(r_*;q)$, and $\pm B(\theta_*;q)$, respectively.
The gray, blue, purple, and brown dashed lines denote $q=q_*$, $q_2$, $q_3$, and $q_4$, respectively. 
Here we set $a=0.9$, and then $r_1^{\mathrm{c}}\simeq 1.56$.
}
\label{fig:ClassIIb_bq_plane}
\end{figure}

In Class~\2-2, there are four characteristic $q$'s: $q_*$, $q_2$, $q_3$, and $q_4$.
They are classified into five cases according to the relative values, which are the same as the cases~\eqref{relative_q_cases_ClassII} of Class~\2 for $a=1$.
The escapable regions are summarized in Table~\ref{table:ClassII} and Fig.~\ref{fig:ClassIIb_bq_plane}.

%%%%%%%%%%%%%%%%%%%%%%%%%%%%%%%%%%%%%%%%%%%%%
%%%%%%%%%%%%%%%%%%%%%%%%%%%%%%%%%%%%%%%%%%%%%

%%%%%%%%%%%%%%%%%%%%%%%%%%%%%%%%%%%%%%%%%%%%%
%%%%%%%%%%%%%%%%%%%%%%%%%%%%%%%%%%%%%%%%%%%%%
\section{Discussions}
\label{sec:summary}
%%%%%%%%%%%%%%%%%%%%%%%%%%%%%%%%%%%%%%%%%%%%%

We have completely classified the necessary and sufficient range of impact parameters $(b,q)$ for photons emitted from the vicinity of a Kerr black hole horizon to escape to infinity, i.e., the escapable regions.
All of the main results are summarized in the tables of Secs.~\ref{sec:classification_ext} and \ref{sec:classification_sub}.
In the process of deriving these results, we have developed a useful method for classification: the visualization of the escapable parameter region in the $b$-$q$ plane.
Furthermore, we have demonstrated a procedure for the systematic identification of the regions.

As mentioned in the Introduction, the evaluation of a photon escape probability is essential to reveal the observability of phenomena in the vicinity of the horizon.
Our complete set of escapable regions provides a basis for evaluating the probability. 
It is worthwhile to comment on further speculations based on the results of our classification.
Naively, we expect the escape probability to decrease as the polar angle of an emitter approaches the poles while the radial coordinate remains fixed because the area of the escapable region simply decreases in this case. 
However, if we focus on photons escaping from the vicinity of the horizon, such naive expectations may not hold because of the effect of the near-horizon geometry. 
In fact, the existence of the critical angles suggests that there is a qualitative difference in the behavior of the escape cone and the escape probability at the critical angles.
In particular, we have identified the special range~$\theta_2\leq \theta_*\leq \pi-\theta_2$ for the extremal case in our context, which universally appears as a region in which the characteristic nature of many phenomena relevant to the spherical photon orbits is found, e.g., the energy extraction efficiency of the collisional Penrose process~\cite{Piran:1975apj,Schnittman:2018ccg} and
high-energy particle collisions near an extremal Kerr horizon~\cite{Banados:2009pr,Harada:2014vka}.
We can expect that within this range the escape probability may be nonzero in the horizon limit because the radial potential barrier exists even in the vicinity of the horizon. 
Even for the fast-spinning (but not extremal) case, since the spherical photon orbits of the horizon class still appear in the vicinity of the horizon~\cite{Igata:2019pgb}, the escape probability may not be zero even if an emitter approaches the horizon.
On the basis of the classification in the present paper, we will report the escape cone and probability of photons emitted from the off-equatorial plane in a forthcoming paper~\cite{Ogasawara:2020prb}.%
\footnote{The escape probability was also discussed in Ref.~\cite{Zulianello:2020cmx}. However, their classification of photon escape does not seem to be complete, although it does represent part of our classification.}

Since the escape probability depends not only on an emitter's position but also its proper motion, we will obtain various nontrivial evaluations by combining our complete set with an emitter's state of interest.
Evaluating such escape probabilities for various states of an emitter is also an important issue for the future.

%%%%%%%%%%%%%%%%%%%%%%%%%%%%%%%%%%%%%%%%%%%%%
\begin{acknowledgments}
%%%%%%%%%%%%%%%%%%%%%%%%%%%%%%%%%%%%%%%%%%%%%
The authors are grateful to Takahiro Tanaka and Kazunori Kohri for useful comments.
This work was supported by
JSPS KAKENHI Grants No.~JP20J00416 and JP20K14467 (K.O.) and No.~JP19K14715 (T.I.).
%%%%%%%%%%%%%%%%%%%%%%%%%%%%%%%%%%%%%%%%%%%%%
\end{acknowledgments}
%%%%%%%%%%%%%%%%%%%%%%%%%%%%%%%%%%%%%%%%%%%%%

\appendix

%%%%%%%%%%%%%%%%%%%%%%%%%%%%%%%%%%%%%%%%%%%%%
%%%%%%%%%%%%%%%%%%%%%%%%%%%%%%%%%%%%%%%%%%%%%
\section{Cases $\theta_*=0$ and $\theta_*=\pi/2$}
\label{App:theta-0-pi}
%%%%%%%%%%%%%%%%%%%%%%%%%%%%%%%%%%%%%%%%%%%%%

We consider photon escape in the case $\theta_*=0$. 
At an emission point on the black hole axis, the regularity of the equations of motion~\eqref{k^theta} and \eqref{eq:kvp} requires $b=0$. This means that all photons escaping from the axis must have zero impact parameter $b$. 
Substituting it into $\Theta(0)\geq 0$, we have $q\geq -a^2$.

Let us focus on the negative range $-a^2\leq q<0$. As shown in Sec.~\ref{subsec:q<0}, the condition $R(r)\geq 0$ gives the allowed range of $q$ in Eq.~\eqref{eq:posiR}. As a result, the following inequality always holds outside the horizon:
\begin{align}
    -a^2\leq q<0<\frac{r}{\Delta} (r^3+a^2r+2a^2).
\end{align}
This implies that all of the photons emitted outwardly with $-a^2\leq q<0$ can escape to infinity.

Let us focus on the non-negative range of $q$. 
Then, two critical values of $q$ appear, 
\begin{align}
    &q_3(0)=q_5(0),
    \\
    &q_2(r_*,0)=q_6(r_*,0)=
    \frac{r_*}{\Delta_*}(r_*^3+a^2r_*+2a^2).
\end{align}
For $a=1$ or for $a<1$ and $r_* \geq r_1^{\mathrm{c}}$, 
together with $q_*$, we can divide these values into three cases,
\begin{align}
\begin{array}{cl}
\text{(i)} & q_*<q_3<q_2,
\\
\text{(\ii)} & q_*=q_3=q_2,
\\
\text{(\iii)} & q_3<q_2<q_*.
\\
\end{array}
\end{align}
The escapable region is summarized in Table~\ref{tab:theta_*=0}.
%%%%%
\begin{table}[t]
    \centering
    \caption{[($a=1$) or ($a<1$ and $r_* \geq r_1^{\mathrm{c}}$)] %\red{
    %\sout{\blue{Escapable region in the $b$-$q$ plane with $\theta_*=0$.}}
    Escapable region $(b,q)$ with $\theta_*=0$.%}
    }
\begin{tabular}{lccc}
\hline\hline
Case & $q$ & $b$ $(\sigma_r=+)$ & $b$ $(\sigma_r=-)$ 
\\ \hline
(i), (\ii)&$q_{\mathrm{min}}\leq q<q_3$&$b=0$&not applicable
\\
&$q_3\leq q\leq 27$&not applicable&not applicable
\\
\hline
(\iii)&$q_{\mathrm{min}}\leq q<q_3$&$b=0$&not applicable
\\
&$q_3\leq q<q_2$&$b=0$&$b=0$
\\
&$q_2\leq q\leq 27$&not applicable&not applicable
\\
\hline\hline
\end{tabular}
    \label{tab:theta_*=0}
\end{table}
%%%%%
On the other hand, for $a<1$ and $r_* < r_1^{\mathrm{c}}$, since $q_*$ does not appear, we only have the single case
\begin{align}
q_3<q_2. 
\end{align}
The escapable region is summarized in Table~\ref{tab:theta_*=0_no2}.

%%%%%
\begin{table}[t]
    \centering
    \caption{($a<1$ and $r_* < r_1^{\mathrm{c}}$) Escapable region $(b,q)$ with $\theta_*=0$.
    }
    \begin{tabular}{ccc}
\hline\hline
$q$ & $b$ $(\sigma_r=+)$ & $b$ $(\sigma_r=-)$ 
\\ \hline
$q_{\mathrm{min}}\leq q<q_3$&$b=0$&not applicable
\\
$q_3\leq q\leq 27$&not applicable&not applicable
\\
\hline\hline
\end{tabular}
    \label{tab:theta_*=0_no2}
\end{table}
%%%%%

~~

We consider photon escape in the case $\theta_*=\pi/2$. 
The non-negativity of $\Theta(\theta_*)$ leads to 
$q\geq 0$. 
Therefore, the necessary parameter regions for a photon to escape to infinity in Table~\ref{table:necessary} of Sec.~\ref{subsec:q>0} are identified with the escapable regions.
The corresponding figures are found in Fig.~\ref{fig:bq_plane}. 
The details of the classification of the escapable region can also be seen in Ref.~\cite{Ogasawara:2019mir}.

\clearpage

%%%%%%%%%%%%%%%%%%%%%%%%%%%%%%%%%%%%%%%%%%%%%
%%%%%%%%%%%%%%%%%%%%%%%%%%%%%%%%%%%%%%%%%%%%%
%%%%%%%%%%%%%%%%%%%%%%%%%%%%%%%%%%%%%%%%%%%%%


\begin{thebibliography}{99}

%
\bibitem{Akiyama:2019cqa}
K.~Akiyama \textit{et al.} (Event Horizon Telescope Collaboration),
First M87 Event Horizon Telescope results. I. The shadow of the supermassive black hole,
Astrophys. J. \textbf{875}, L1 (2019)
%doi:10.3847/2041-8213/ab0ec7
[arXiv:1906.11238 [astro-ph.GA]].

%
\bibitem{Cardoso:2019rvt}
V.~Cardoso and P.~Pani,
Testing the nature of dark compact objects: A status report,
Living Rev. Relativity \textbf{22}, 4 (2019)
%doi:10.1007/s41114-019-0020-4
[arXiv:1904.05363 [gr-qc]].

%
\bibitem{Synge:1966okc}
J.~L.~Synge,
The escape of photons from gravitationally intense stars,
Mon. Not. R. Astron. Soc. \textbf{131}, 463 (1966).
%doi:10.1093/mnras/131.3.463

%
\bibitem{Semerak:1996}
O.~Semerak, 
Photon escape cones in the Kerr field, 
Helv. Phys. Acta \textbf{69}, 69 (1996).

%

\bibitem{Stuchlik:2018qyz}
Z.~Stuchl\'\i{}k, D.~Charbul\'ak, and J.~Schee,
Light escape cones in local reference frames of Kerr-de Sitter black hole spacetimes and related black hole shadows,
Eur. Phys. J. C \textbf{78}, 180 (2018)
%doi:10.1140/epjc/s10052-018-5578-6
[arXiv:1811.00072 [gr-qc]].


%
\bibitem{Ogasawara:2016yfk}
K.~Ogasawara, T.~Harada, U.~Miyamoto, and T.~Igata,
Escape probability of the super-Penrose process,
Phys. Rev. D \textbf{95}, 124019 (2017)
%doi:10.1103/PhysRevD.95.124019
[arXiv:1609.03022 [gr-qc]].

%
\bibitem{Ogasawara:2019mir} 
  K.~Ogasawara, T.~Igata, T.~Harada, and U.~Miyamoto,
  Escape probability of a photon emitted near the black hole horizon,
  Phys.\ Rev.\ D \textbf{101}, 044023 (2020)
%doi:10.1103/PhysRevD.101.044023
  [arXiv:1910.01528 [gr-qc]].
  
%
\bibitem{Igata:2019hkz}
T.~Igata, K.~Nakashi, and K.~Ogasawara,
Observability of the innermost stable circular orbit in a near-extremal Kerr black hole,
Phys.\ Rev.\ D \textbf{101}, 044044 (2020)
%doi:10.1103/PhysRevD.101.044044
[arXiv:1910.12682 [astro-ph.HE]].

%
\bibitem{Gates:2020els}
D.~E.~A.~Gates, S.~Hadar, and A.~Lupsasca,
Photon emission from circular equatorial Kerr orbiters,
Phys. Rev. D \textbf{103}, 044050 (2021)
%doi:10.1103/PhysRevD.103.044050
[arXiv:2010.07330 [gr-qc]].


%
\bibitem{Gates:2020sdh}
D.~E.~A.~Gates, S.~Hadar, and A.~Lupsasca,
Maximum observable blueshift from circular equatorial Kerr orbiters,
Phys. Rev. D \textbf{102}, no.10, 104041 (2020)
%doi:10.1103/PhysRevD.102.104041
[arXiv:2009.03310 [gr-qc]].


%
\bibitem{Zhang:2020pay}
M.~Zhang and J.~Jiang,
Escape probability of a photon near the horizon of Kerr-Sen black hole,
Phys. Rev. D \textbf{102}, no.12, 124012 (2020)
%doi:10.1103/PhysRevD.102.124012
[arXiv:2004.11087 [gr-qc]].

%
\bibitem{Bardeen:1972}
J.~M.~Bardeen, W.~H.~Press, and S.~A.~Teukolsky,
Rotating black holes: Locally nonrotating frames, energy extraction, and scalar synchrotron radiation, 
Astrophys.\ J.\ \textbf{178}, 347 (1972).

%
\bibitem{Bardeen:1973tla}
J.~M.~Bardeen,
Timelike and null geodesics in the Kerr metric,
in \textit{Black Holes (Les Astres Occlus)}, edited by C.~Dewitt and B.~S.~Dewitt (Gordon and Breach, NewYork, 1973), pp. 215--239.

%
\bibitem{Igata:2019pgb}
T.~Igata, H.~Ishihara, and Y.~Yasunishi,
Observability of spherical photon orbits in near-extremal Kerr black holes,
Phys.\ Rev.\ D \textbf{100}, 044058 (2019)
%doi:10.1103/PhysRevD.100.044058
[arXiv:1904.00271 [gr-qc]].


%
\bibitem{Walker:1970un}
M.~Walker and R.~Penrose,
On quadratic first integrals of the geodesic equations for type \{22\} spacetimes,
Commun. Math. Phys. \textbf{18}, 265 (1970).
%doi:10.1007/BF01649445

%
\bibitem{Carter:1963}
B.~Carter,
Global structure of the Kerr family of gravitational fields, 
Phys.\ Rev.\ \textbf{174}, 1559 (1968).

%
\bibitem{Teo:2003}
E. Teo,
Spherical photon orbits around a Kerr black hole,
Gen.\ Relativ.\ Gravit.\ \textbf{35}, 1909 (2003).

%
\bibitem{Piran:1975apj}
T.~Piran, J.~Shaham, and J.~Katz,
High efficiency of the Penrose mechanism for particle collisions,
Astrophys.\ J.\ \textbf{196}, L107 (1975).

%
\bibitem{Schnittman:2018ccg}
J.~D.~Schnittman,
The collisional Penrose process,
Gen. Relativ. Gravit. \textbf{50}, 77 (2018)
%doi:10.1007/s10714-018-2373-5
[arXiv:1910.02800 [astro-ph.HE]].

%
\bibitem{Banados:2009pr}
M.~Banados, J.~Silk, and S.~M.~West,
Kerr Black Holes as Particle Accelerators to Arbitrarily High Energy,
Phys. Rev. Lett. \textbf{103}, 111102 (2009)
%doi:10.1103/PhysRevLett.103.111102
[arXiv:0909.0169 [hep-ph]].

%
\bibitem{Harada:2014vka}
T.~Harada and M.~Kimura,
Black holes as particle accelerators: A brief review,
Classical Quantum Gravity \textbf{31}, 243001 (2014)
%doi:10.1088/0264-9381/31/24/243001
[arXiv:1409.7502 [gr-qc]].

%
\bibitem{Ogasawara:2020prb}
K.~Ogasawara and T.~Igata (to be published).

%
\bibitem{Zulianello:2020cmx}
A.~Zulianello, R.~Carballo-Rubio, S.~Liberati, and S.~Ansoldi,
Electromagnetic tests of horizonless rotating black hole mimickers,
arXiv:2005.01837 [gr-qc].

\end{thebibliography}
\end{document}